\documentclass{jfm}
\usepackage{graphicx}

\usepackage{color}
\usepackage[dvipsnames]{xcolor}

\usepackage{amsmath,amssymb,amsfonts,bm}
\usepackage{mathtools}

\usepackage[english]{babel}
\usepackage[utf8]{inputenc} \label{key}
\usepackage[T1]{fontenc}

\usepackage{hyperref}
\hypersetup{
     colorlinks=true,
    linkcolor=blue,
    filecolor=magenta,      
    urlcolor=blue,
    citecolor=blue,
}

\newcommand{\DiracDelta}[1]{\operatorname{\delta}\left(#1\right)}

\newcommand{\vect}[1]{\boldsymbol{#1}}

\newcommand{\R}{\vect{r}}

\newcommand{\Intd}{\mathrm{d}}

\newcommand{\halb}{\frac{1}{2}}

\definecolor{AirForceBlue}{rgb}{0., 0.34, 0.66}

\shorttitle{Axisymmetric Stokeslet between two disks}
\shortauthor{A. Daddi-Moussa-Ider, A. R. Sprenger, Y. Amarouchene, T. Salez, et al.}
\title{Axisymmetric Stokes flow due to a point-force singularity acting between two coaxially positioned rigid no-slip disks}

\author{Abdallah Daddi-Moussa-Ider\aff{1}
\corresp{\email{abdallah.daddi.moussa.ider@uni-duesseldorf.de}}, 
Alexander R. Sprenger\aff{1}, \\
Yacine Amarouchene\aff{2}, Thomas Salez\aff{2,3}, Clarissa Sch\"{o}necker \aff{4,5}, Thomas Richter\aff{6}, Hartmut L\"{o}wen\aff{1}, and Andreas M. Menzel\aff{1}
} 

\affiliation{
\aff{1}Institut f\"{u}r Theoretische Physik II: Weiche Materie, Heinrich-Heine-Universit\"{a}t D\"{u}sseldorf, D-40225 D\"{u}sseldorf, Germany
\aff{2}Univ. Bordeaux, CNRS, LOMA, UMR 5798, F-33405, Talence, France
\aff{3}Global Station for Soft Matter, Global Institution for Collaborative Research and Education, Hokkaido University, Sapporo, Hokkaido 060-0808, Japan
\aff{4} Technische  Universit\"{a}t  Kaiserslautern,  D-67663  Kaiserslautern,  Germany
\aff{5} Max-Planck-Institut f\"{u}r Polymerforschung,  D-55218  Mainz,  Germany
\aff{6} Institut für Analysis und Numerik, Otto-von-Guericke-Universität Magdeburg, D-39106 Magdeburg, Germany
}

\usepackage{upgreek}

\usepackage{nicefrac}

\begin{document}

\maketitle

\begin{abstract}
	We investigate theoretically on the basis of the steady Stokes equations for a viscous incompressible fluid the flow induced by a Stokeslet located on the centre axis of two coaxially positioned rigid disks.
	The Stokeslet is directed along the centre axis. 
	No-slip boundary conditions are assumed to hold at the surfaces of the disks.
	We perform the calculation of the associated Green's function  in large parts analytically, reducing the spatial evaluation of the flow field to one-dimensional integrations amenable to numerical treatment.
	To this end, we formulate the solution of the hydrodynamic problem for the viscous flow surrounding the two disks as a mixed-boundary-value problem, which we then reduce into a system of four dual integral equations.  
	We show the existence of viscous toroidal eddies arising in the fluid domain bounded by the two disks, manifested in the plane containing the centre axis through adjacent counterrotating eddies.
	Additionally, we probe the effect of the confining disks on the slow dynamics of a point-like particle by evaluating the hydrodynamic mobility function associated with axial motion.
	Thereupon, we assess the appropriateness of the commonly-employed superposition approximation and discuss its validity and applicability as a function of the geometrical properties of the system.  
	Additionally, we complement our semi-analytical approach by finite-element computer simulations, which reveals a good agreement.
	Our results may find applications in guiding the design of microparticle-based sensing devices and electrokinetic transport in small scale capacitors.
\end{abstract}



\section{Introduction}
\label{sec:intro}

Manipulating colloidal particles suspended in viscous media is a challenging task and is of paramount importance in various fields of engineering and natural sciences.
Frequently, taking into account the fluid-mediated hydrodynamic interactions between particles moving through a liquid is essential to predict the behaviour of colloidal suspensions and polymer solutions~\citep{probstein05, mewis12}.
Recent advances in micro- and nanofluidic technologies have permitted the fabrication and manufacturing of channels with well-defined geometries and characteristic dimensions ranging from the micro- to the nanoscale.
A deep understanding of the nature of the mutual interactions between particles and their confining interfaces is of crucial importance in guiding the design of devices and tools for an optimal nanoscale control of biological macromolecules.
Notable examples include single-molecule manipulation~\citep{turner98, campbell04}, DNA mapping for genomic applications~\citep{riehn05, reisner05, persson10}, DNA separation and sorting~\citep{doyle02, cross07, xia12}, and rheological probing of complex structures using Atomic Force Microscopy cantilevers~\citep{franccois08, franccois09, dufour12, darwiche13}.

At these small scales, fluid flows are governed by low-Reynolds-number hydrodynamics, where viscous effects dominate over inertial effects~\citep{kim13}.
Solutions for fluid flows due to point forces, or Stokeslets, acting close to confining boundaries have been tabulated for various types of geometries, as summarised in the classic textbook by \cite{happel12}.
The study of the fluid-mediated hydrodynamic interactions in a channel confinement has received significant attention from many researchers over the past couple of years.
In the following, we provide a survey of the current state of the art and summarise the relevant literature in this subject.

The first attempt to address the motion of a spherical particle confined between two infinitely extended no-slip walls dates back to \cite{faxen21}, who calculated in his PhD dissertation the hydrodynamic mobility parallel to the walls.
These calculations have been performed when the particle is located in the quarter-plane or mid-plane between the two confining walls~\citep{happel12}.
Later, \cite{oseen28} suggested that the hydrodynamic mobility between two walls could approximately be obtained by superposition of the contributions resulting from each single wall. 
A modified coherent superposition approximation has further been suggested by~\cite{benesch03}, providing the diffusion coefficients of a Brownian sphere in confining channels.
These predictions were found to match more accurately the existing experimental data reported in the literature.

Exact solutions for a point-force singularity acting at an arbitrary position between two walls have first been obtained using the image technique in a seminal article by \cite{liron76}.
It has been noted that the effect of the second wall becomes important when the distance separating the particle from the closest wall is larger than approximately one tenth of the channel width~\citep{brenner99}.
Using this solution, \cite{liron78cilia} further investigated the fluid transport problem of cilia between two parallel plates.
A joint analytical-numerical approach~\citep{ganatos80b, ganatos80a} as well as a multipole expansion technique~\citep{swan10} were presented to address the motion of an extended particle confined between two hard walls.
Meanwhile, \cite{bhattacharya02} constructed the image system for the flow field produced by a force multipole in a space bounded by two parallel walls using the image representation for Stokes flow.
In addition, compressibility effects were examined by \cite{felderhof06twoWalls, felderhof10echoing, felderhof10loss}.
In this context, \cite{hackborn90} investigated the asymmetric Stokes flow between two parallel planes due to a rotlet singularity, the axis of which is parallel to the boundary planes. 
Further, \cite{ozarkar08} prescribed an analytical approach using the image-system technique for determining the Stokes flow around particles in a thin film bounded by a wall and a gas-liquid interface.
More recently, \cite{daddi16b} have provided the frequency-dependent hydrodynamic mobility functions between two planar elastic interfaces endowed with resistance toward shear and bending deformation modes.

Experimentally, \cite{dufresne01} reported direct imaging measurements of a colloidal particle diffusing between two parallel surfaces, finding a good agreement with the superposition approximation suggested by Oseen.
In addition, video microscopy~\citep{faucheux94} combined with optical tweezers~\citep{lin00, traenkle16} as well as dynamic light scattering~\citep{lobry96} have also allowed for good agreement with available theoretical predictions.
Further experimental investigations have focused on DNA conformation and diffusion in slit-like confinements~\citep{stein06, balducci06, strychalski08,tang10, graham11, dai13, jones13}.

Concerning collective properties, the behaviour of suspensions in a channel bounded by two planar walls has received a lot of attention.
For instance, \cite{bhattacharya05b} examined the fluid-mediated hydrodynamic interactions in a suspension of spherical particles confined between two parallel planar walls under creeping-flow conditions.
In addition, \cite{bhattacharya08} considered the collective motion of a two-dimensional periodic array of colloidal particles in a slit pore. 
Using a novel accelerated Stokesian-dynamics algorithm, \cite{baron08} performed fully-resolved computer simulations to investigate the collective motion of linear trains and regular square arrays of particles suspended in a viscous fluid bounded by two parallel plates.
Further, \cite{blaw08} analysed the far-field response to external forcing of a suspension of particles in a channel.
Meanwhile, \cite{swan11} presented a numerical method for computing the hydrodynamic forces exerted on particles in a suspension confined between two parallel walls.
Furthermore, \cite{saintillan06} employed Brownian dynamics simulations to investigate the effect of chain flexibility on the cross-streamline migration of short polymers in a pressure-driven flow between two flat plates.
The latter numerical study confirmed the existence of a shear-induced migration toward the channel centreline away from the confining solid  boundaries.

The hydrodynamic problem of particles freely moving between plane-parallel walls in the presence of an incident flow has further been considered in still more details.
Under an external flow, \cite{uspal13} showed how shape and geometric confinement of rigid microparticles can conveniently be tailored for self-steering.
\cite{jones04} made use of a two-dimensional Fourier-transform technique to obtain an analytic expression of the Green tensor for the Stokes equations with an incident Poiseuille flow. 
In addition, he provided the elements of the resistance and mobility tensors in this slit-like geometry.
Besides, \cite{bhattacharya06pof} introduced a novel numerical algorithm based on transformations between Cartesian and spherical representations of Stokes flow to account for an incident Poiseuille flow.
Meanwhile, \cite{staben03} presented a novel boundary-integral algorithm for the motion of a particle between two parallel planar walls in Poiseuille flow.
The boundary-integral method formulated in their work allowed to directly incorporate the effects of the confining walls into the stress tensor, without requiring discretisation of the two walls.
In this context, \cite{griggs07} and \cite{janssen07, janssen08} employed boundary-integral methods to examine the motion of a deformable drop between two parallel walls in Poiseuille flow, where lateral migration towards the channel centre is observed.

Geometric confinements significantly alter the behaviour of swimming microorganisms and can affect the motility of self-propelling active particles in a pronounced way~\citep{lauga09, menzel13, menzel15, lauga2016ARFM, zottl16, bechinger16, ostapenko2018, gompper20, shaebani20}.
Surface-related effects on microswimmers can lead to crucial implications for biofilm formation and microbial activity.
In a channel bounded by two walls, \cite{bilbao13} studied the locomotion of a model nematode, finding that the swimming organism tends to swim faster and navigate more effectively under confinement. 
Furthermore, \cite{wu15, wu16} investigated the effect of confinement on the swimming behaviour of a model eukaryotic cell undergoing amoeboid motion.
There, the swimmer has been modeled as an inextensible membrane deploying local active force.
It has been found that confinement can strongly alter the swimming gait.
In addition, \cite{brotto13} described theoretically the dynamics of self-propelling active particles in rigidly confined thin liquid films.
They demonstrated that, due to hydrodynamic friction with the nearby rigid walls, confined microswimmers do not only reorient themselves in response to flow gradients but they can also show reorientation in uniform flows.
In this context, \cite{mathijssen2015hydrodynamics} investigated theoretically the hydrodynamics of self-propelling microswimmers in a thin film.
Besides, \cite{daddi2018state} examined the behaviour of a three-sphere microswimmer in a channel bounded by two walls, where different swimming states have been observed.
More recently, amoeboid swimming in a compliant channel was numerically investigated~\citep{dalal20}.

In all of the above-mentioned studies, the confining channel was assumed to be of infinite extent or periodically replicated along the lateral directions.
Instead, we here consider the hydrodynamic problem for a point force acting near two coaxially positioned disks of finite radius.
In many biologically and industrially relevant applications, finite-size effects become crucial for an accurate and reliable description of transport processes ranging from the microscale to the nanoscale.
Prime examples include the ionic transport and electrokinetics in small scale capacitors~\citep{marini12, thakore15, babel18, asta19}, electrochemomechanical energy conversion in microfluidic channels~\citep{daiguji04}, and the rheology of droplets, capsules, or cells in constricted/structured microchannels~\citep{park13, legoff17, tregouet18, tregouet19}, where boundary effects may play a pivotal role.

In this contribution, we take a step toward addressing this context by presenting an analytical theory for the viscous flow resulting from a Stokeslet singularity acting along the centre axis of two coaxially positioned disks of no-slip surfaces.
We formulate the hydrodynamic problem as a mixed-boundary-value problem, which we then transform into a system of dual integral equations.
Along this path,  we show that the solution of the flow field in the fluid region bounded by the two disks exhibits viscous toroidal eddies.
In addition to that, we derive expressions for the hydrodynamics mobility functions and discuss the applicability and limitations of the superposition approximation. 
Moreover, we support our semi-analytical results by numerical simulations using a finite-element method (FEM), which leads to a good agreement.

The remainder of this article is organised as follows.
In Sec.~\ref{sec:mathematik}, we formulate the problem mathematically and derive the corresponding system of dual integral equations, from which the solution for the hydrodynamic flow fields can be obtained.
We then make use of this solution in Sec.~\ref{sec:mobilitaet} to yield an integral expression of the mobility function of a point-like particle slowly translating along the axis of the disks.
Concluding remarks and outlooks are contained in Sec.~\ref{sec:conclusions}.
In Appendix~\ref{Appendix}, we detail the analytical derivation of the kernel functions arising in the resulting integral equations.

\section{Mathematical formulation}
\label{sec:mathematik}

\begin{figure}
	\centering
	\includegraphics[scale=1.2]{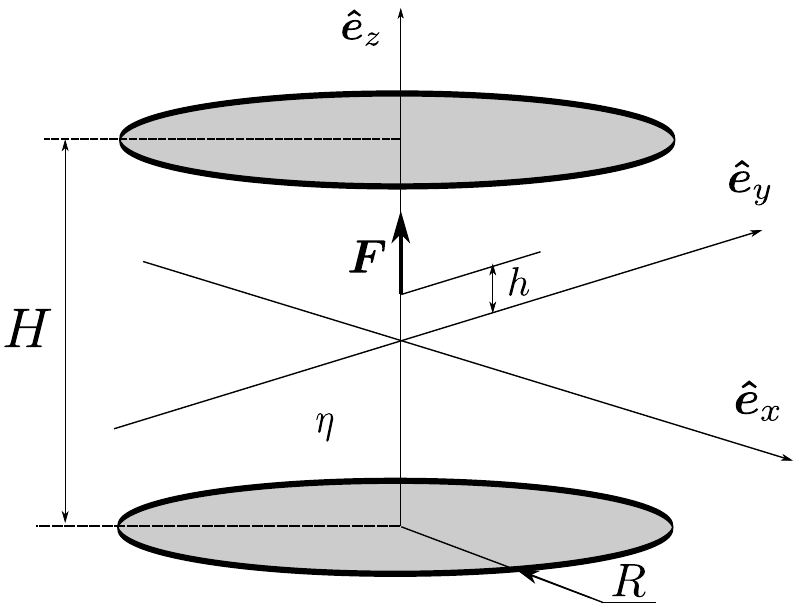}
	\caption{Schematic of the system.
	The surrounding viscous Newtonian fluid is set into motion through the action of a point-force singularity located on the axis of the symmetry axis of two coaxially positioned disks.}
	\label{sketch}
\end{figure}

We examine the axisymmetric flow induced by a Stokeslet singularity acting on the axis of two coaxially positioned circular disks of equal radius~$R$.
Moreover, we suppose that the disks are located within the planes~$z=-H/2$ and $z=H/2$ with~$H$ denoting the separation distance between the disks. Their centres are positioned on the $z$ axis.
In addition, we assume that the surrounding viscous fluid is Newtonian, of constant dynamic viscosity~$\eta$, and that the flow is incompressible.

\subsection{Governing equations}

In low-Reynolds-number hydrodynamics, the fluid dynamics is governed by the Stokes equations~\citep{happel12}
\begin{subequations}\label{StokesGleischungen}
	\begin{align}
		\boldsymbol{\nabla} \boldsymbol{\cdot} \vect{v} &= 0 \, , \\
		\boldsymbol{\nabla} \boldsymbol{\cdot} \boldsymbol{\sigma} + 
		F \delta(\R-\R_0) \, \vect{\hat{e}}_z  &= \vect{0} \, , 
	\end{align}
\end{subequations}
where~$\vect{v}$ and $\boldsymbol{\sigma}$ denote, respectively, the fluid velocity field and the hydrodynamic stress tensor.
For a Newtonian fluid, the latter is given by $\boldsymbol{\sigma} = -p \vect{I} + 2\eta \vect{E}$, where~$p$ is the pressure field and $\vect{E} = \tfrac{1}{2} \left( \boldsymbol{\nabla} \vect{v} + (\boldsymbol{\nabla} \vect{v})^\mathrm{T} \right)$ is the rate-of-strain tensor, with the superscript~T denoting a transpose.
In addition, $\delta$ stands for the Dirac delta function, and $F$ is the amplitude of a stationary point force acting on the fluid at position~$\R_0 = h \vect{\hat{e}}_z$, where $-H/2 < h < H/2$, with~$\vect{\hat{e}}_z$ denoting the unit vector along the $z$~direction.
See Fig.~\ref{sketch} for an illustration of the system setup.
In the remainder of this article, we scale all the lengths involved in the problem by the separation~$H$ of the two disks.

We designate by the subscript~1 the variables and parameters in the fluid region underneath the plane containing the lower disk, for which $z \le -1/2$, by the subscript~2 the fluid domain bounded by the planes $z=-1/2$ and~$z=1/2$, and by the subscript~3 the region above the plane containing the upper disk, for which $z \ge 1/2$.
Since the system is axisymmetric, all field variables are thus functions of the radial and axial coordinates only.
Accordingly, the Stokes equations~\eqref{StokesGleischungen} can be projected onto the cylindrical coordinate system as
\begin{subequations}\label{StokesGleischungenCylind}
	\begin{align}
		\frac{v_r}{r} + \frac{\partial v_r}{\partial r} + \frac{\partial v_z}{\partial z} &= 0 \, , \\
		-\frac{\partial p}{\partial r} + \eta \left( \Updelta v_r - \frac{v_r}{r^2} \right) &= 0 \, , \\
		-\frac{\partial p}{\partial z} + \eta \Updelta v_z + F \delta(\R - \R_0) &= 0 \, ,
	\end{align}
\end{subequations}
wherein~$v_r$ and~$v_z$ denote the radial and axial fluid velocities, respectively, and~$\Updelta$ is the Laplace operator given by
\begin{equation}
	\Updelta := \frac{\partial^2}{\partial r^2} + \frac{1}{r} \frac{\partial}{\partial r} + \frac{\partial^2}{\partial z^2} \, .
\end{equation}
We note that the three-dimensional Dirac delta function is expressed in axisymmetric cylindrical coordinates as $\delta(\R - \R_0) = (\pi r)^{-1} \delta(r)\delta(z-h)$ \citep{bracewell99}.

In an unbounded viscous fluid, \textit{i.e.}, in the absence of the disks, the solution of Eqs.~\eqref{StokesGleischungenCylind} is given by the Oseen tensor, commonly denominated as the free-space Green function~\citep{kim13}
\begin{equation}
	v_r^\mathrm{S} = \frac{F}{8\pi\eta} \frac{r \left(z-h\right)}{\rho^3} \, , \qquad 
	v_z^\mathrm{S} = \frac{F}{8\pi\eta} \left( \frac{2}{\rho} - \frac{r^2}{\rho^3} \right) \, , 
	\label{VS}
\end{equation}
with the distance from the position of the point force $\rho = \left(r^2 + \left(z-h\right)^2\right)^{\nicefrac{1}{2}}$.
The corresponding pressure field reads
\begin{equation}
	p^\mathrm{S} = \frac{F}{4\pi} \frac{z-h}{\rho^3} \, . \label{pS}
\end{equation}

In the presence of the confining disks, the solution of the flow problem can be expressed as a superposition of the solution in an unbounded fluid, given above by Eqs.~\eqref{VS} and~\eqref{pS}, and a complementary solution, the sum of the two solutions being required to satisfy the underlying regularity and boundary conditions.
Then
\begin{equation}
	\vect{v} = \vect{v}^\mathrm{S} + \vect{v}^* \, , \qquad
	p = p^\mathrm{S} + p^* \, ,  
\end{equation}
wherein~$\vect{v}^*$ and~$p^*$ stand for the complementary solutions (also referred to as the image solution~\citep{ blake71}) for the velocity and pressure fields, respectively.

For an axisymmetric Stokes flow, the general solution can be expressed in terms of two harmonic functions $\phi$ and~$\psi$ as~\citep{imai73, kim83}
\begin{equation}
	v_r^* = z \, \frac{\partial \phi}{\partial r} + \frac{\partial \psi}{\partial r} \, , \qquad
	v_z^* = z \, \frac{\partial \phi}{\partial z} - \phi + \frac{\partial \psi}{\partial z} \, , \qquad
	p^* = 2\eta \, \frac{\partial \phi}{\partial z} \, ,
\end{equation}
with
\begin{equation}
	\Updelta \phi = 0 \, , \qquad \Updelta \psi = 0 \, . \label{laplacePhiUndPsi}
\end{equation}

In each of the three fluid domains introduced above, the solution of Laplace's Eqs.~\eqref{laplacePhiUndPsi} can be expressed in terms of Fourier-Bessel integrals as 
\begin{subequations}
	\begin{align}
		\phi_i &= \frac{F}{8\pi\eta} \int_0^\infty \left( A_i^+(\lambda) e^{\lambda z} + A_i^-(\lambda) e^{-\lambda z} \right) J_0 (\lambda r) \, \Intd \lambda \, , \\
		\psi_i &= \frac{F}{8\pi\eta} \int_0^\infty \left( B_i^+(\lambda) e^{\lambda z} + B_i^-(\lambda) e^{-\lambda z} \right) J_0 (\lambda r) \, \Intd \lambda \, ,
	\end{align}
\end{subequations}
for~$i \in \{1,2,3\}$, with~$\lambda$ denoting the wavenumber and $J_k$ the $k$th-order Bessel function of the first kind~\citep{abramowitz72}.
In addition, $A_i^\pm$ and~$B_i^\pm$ are wavenumber-dependent unknown coefficients, to be determined from the regularity and boundary conditions.
Then, the components of the image velocity and pressure fields are given by
\begin{subequations}\label{ExpressionsVelocityPressure}
	\begin{align}
		{v_r^*}_i &= - \frac{F}{8\pi\eta} \int_0^\infty \lambda \left( \left( z A_i^+ + B_i^+ \right) e^{\lambda z}
		 + \left( z A_i^- + B_i^- \right) e^{-\lambda z} \right) J_1(\lambda r) \, \Intd \lambda \, , \\
		 {v_z^*}_i &= - \frac{F}{8\pi\eta} \int_0^\infty \left( E_i^+e^{\lambda z} + E_i^- e^{-\lambda z} \right) J_0(\lambda r) \, \Intd \lambda \, , \\
		 p^*_i &= \frac{F}{4\pi} \int_0^\infty \lambda \left( A_i^+ e^{\lambda z} - A_i^- e^{-\lambda z} \right) J_0(\lambda r) \, \Intd \lambda \, ,
	\end{align}
\end{subequations}
for $i \in \{1,2,3\}$, where we have defined the abbreviations $E_i^\pm = \left( 1 \mp \lambda z \right) A_i^\pm \mp \lambda B_i^\pm$.

\subsection{Boundary conditions and dual integral equations}

As regularity conditions, we require for the image field vanishing velocity and pressure far away from the singularity location as~$\rho\to\infty$.
This implies that~$A_1^- = B_1^- = A_3^+ = B_3^+ = 0$.
In what follows, to simplify notations, we drop the plus sign in the fluid domain underneath the lower disk to denote $A_1 = A_1^+$ and $B_1 = B_1^+$, and we drop the minus sign in the fluid domain above the upper disk to denote $A_3 = A_3^-$ and $B_3 = B_3^-$.

The boundary conditions consist of requiring (a)~the natural continuity of the total fluid velocity field at the interfaces between the fluid domains, 
(b)~vanishing total velocities at the surfaces of the disks (the no-slip and no-permeability boundary condition~\citep{lauga07noslip}),
and~(c)~continuity of the total viscous-stress vectors at the interfaces between the fluid domains outside the regions occupied by the disks.
Mathematically, these conditions can be expressed as
\begin{subequations}
	\begin{align}
		 (\vect{v}_1 - \vect{v}_2) \rvert_{z=-\nicefrac{1}{2}} \,=\, (\vect{v}_2 - \vect{v}_3) \rvert_{z=\nicefrac{1}{2}} &= \vect{0} \qquad\quad (r > 0) \, , \label{KontinEq} \\
		 \vect{v}_1 \rvert_{z = -\nicefrac{1}{2}} \,=\, \vect{v}_2 \rvert_{z = \pm \nicefrac{1}{2}} \,=\, \vect{v}_3 \rvert_{z = \nicefrac{1}{2}} &= \vect{0} 
		 \qquad\quad (r<R) \, , \label{InnerProblemOrig} \\
		 \left( \boldsymbol{\sigma}_2 - \boldsymbol{\sigma}_1 \right) \boldsymbol{\cdot} \vect{\hat{e}}_z \rvert_{z=-\nicefrac{1}{2}} \,=\,
		 \left( \boldsymbol{\sigma}_3 - \boldsymbol{\sigma}_2 \right) \boldsymbol{\cdot} \vect{\hat{e}}_z \rvert_{z=\nicefrac{1}{2}} &= \vect{0}	
		 \qquad\quad (r>R) \, , \label{OuterProblemOrig}
	\end{align}
\end{subequations}
where the components of the stress vector are expressed in cylindrical coordinates for an axisymmetric flow field by
\begin{equation}
	\boldsymbol{\sigma}_i \boldsymbol{\cdot} \vect{\hat{e}}_z =
	\eta \left( \frac{\partial {v_r}_i}{\partial z} + \frac{\partial {v_z}_i}{\partial r} \right) \vect{\hat{e}}_r
	+ \left( -p_i + 2\eta \, \frac{\partial {v_z}_i}{\partial z} \right) \vect{\hat{e}}_z \ , \qquad
	i \in \{1,2,3\} \, .
\end{equation}

Applying the continuity of the radial components of the fluid velocity at the surfaces occupied by the two disks yields the expressions of the wavenumber-dependent coefficients associated with the intermediate fluid domain bounded by the two disks as functions of those in the lower and upper fluid domains.
Defining $\vect{X}_2 = \left( A_2^-, B_2^-, A_2^+, B_2^+ \right)^\mathrm{T}$ and $\vect{X}_{13} = \left( A_1, B_1, A_3, B_3 \right)^\mathrm{T}$, we obtain
\begin{equation}
	\vect{X}_2 = \vect{Q} \boldsymbol{\cdot} \vect{X}_{13} \, , \label{X2}
\end{equation}
where the matrix~$\vect{Q}$ is given by 
\begingroup
\renewcommand*{\arraystretch}{1.4}
\begin{equation}\label{Q}
	\vect{Q} = \left( s^2-\lambda^2 \right)^{-1}
	\begin{pmatrix}
		\tfrac{1}{2} \left(s+\lambda c\right) & - \lambda s & - \tfrac{1}{2} \, \phi^+ & - \lambda^2 \\
		\tfrac{1}{4} \lambda s & \tfrac{1}{2} \left(s-\lambda c\right) & -\tfrac{1}{4} \lambda^2 & - \tfrac{1}{2} \, \phi^- \\
		- \tfrac{1}{2} \, \phi^+ & \lambda^2 & \tfrac{1}{2} \left(s+\lambda c\right) & \lambda s \\
		\tfrac{1}{4} \lambda^2 & - \tfrac{1}{2} \, \phi^- & -\tfrac{1}{4} \lambda s & \tfrac{1}{2} \left( s-\lambda c \right)
	\end{pmatrix} \, .
\end{equation}
\endgroup
Here, we have defined for convenience the abbreviations~$s = \sinh (\lambda)$ and $c = \cosh(\lambda)$.
In addition, $\phi^\pm = \lambda \left(\lambda \pm 1\right) + se^{-\lambda}$.

On the one hand, by addressing the no-slip velocity boundary conditions at the surfaces of the disks prescribed by Eqs.~\eqref{InnerProblemOrig} and projecting the resulting equations onto the radial and tangential directions, four integral equations on the \emph{inner} domain are obtained,
\begin{subequations}\label{InnerProblem}
	\begin{align}
		\int_0^\infty \lambda \left( \tfrac{1}{2} A_1 - B_1 \right) e^{-\tfrac{\lambda}{2}} J_1(\lambda r) \, \Intd \lambda &= \psi_1^+ (r) \qquad (r<R) \, , \\
		\int_0^\infty \lambda \left(\tfrac{1}{2} A_3 + B_3 \right) e^{-\tfrac{\lambda}{2}} J_1(\lambda r) \, \Intd \lambda &= \psi_1^- (r) \qquad (r<R) \, , \\
		\int_0^\infty \left( A_1 + \lambda\left(\tfrac{1}{2}A_1-B_1\right) \right) e^{-\tfrac{\lambda}{2}} J_0(\lambda r) \, \Intd \lambda &= \psi_2^+ (r) \qquad (r<R) \, , \\
		\int_0^\infty \left( A_3 + \lambda\left(\tfrac{1}{2}A_3+B_3\right) \right) e^{-\tfrac{\lambda}{2}} J_0(\lambda r) \, \Intd \lambda &= \psi_2^- (r) \qquad (r<R) \, ,
	\end{align}
\end{subequations}
wherein the terms appearing on the right-hand sides in these equations are radial functions resulting from the evaluation of the terms associated with the flow velocity field induced by the free-space Stokeslet at the surfaces of the coaxially positioned disks. They are explicitly given by
\begin{equation}\label{psi}
	\psi_1^\pm (r) = \frac{\pm r \left( h \pm \frac{1}{2} \right)}{\left(r^2+\left(h \pm\frac{1}{2}\right)^2\right)^{\frac{3}{2}}} \, , \qquad
	\psi_2^\pm (r) = \frac{ r^2 + 2 \left(h\pm \frac{1}{2}\right)^2}{\left(r^2+\left(h\pm \frac{1}{2}\right)^2\right)^{\frac{3}{2}}} \, .
\end{equation}

On the other hand, four integral equations on the \emph{outer} domain are obtained by addressing the continuity of the hydrodynamic stress vector at $z=\pm 1/2$ prescribed by Eq.~\eqref{OuterProblemOrig}.
They can be cast in the form
\begin{subequations}\label{OuterProblem}
	\begin{align}
		\int_0^\infty g_i(\lambda) J_1(\lambda r) \, \Intd \lambda &= 0 \qquad (r>R) \, , \quad i \in \{1,3\} \, , \label{outerProblem13} \\
		\int_0^\infty g_i(\lambda) J_0(\lambda r) \, \Intd \lambda &= 0 \qquad (r>R) \, , \quad i \in \{2,4\} \, , \label{outerProblem24}
	\end{align}
\end{subequations}
where we have defined the wavenumber-dependent quantities
\begin{subequations}\label{g1234}
	\begin{align}
		g_1(\lambda) &= \lambda^2 \left( \left( \tfrac{1}{2} A_2^--B_2^- \right)e^{\tfrac{\lambda}{2}} + \left( \tfrac{1}{2} \left( A_1 - A_2^+ \right) + B_2^+ - B_1 \right) e^{-\tfrac{\lambda}{2}} \right) , \\
		g_3(\lambda) &= \lambda^2 \left( \left( \tfrac{1}{2} A_2^+ + B_2^+ \right) e^{\tfrac{\lambda}{2}} + \left( \tfrac{1}{2} \left( A_3 - A_2^- \right) + B_3-B_2^- \right) e^{-\tfrac{\lambda}{2}} \right) , \\
		g_2(\lambda) &=  C^- e^{\tfrac{\lambda}{2}} + \lambda \left( \left(1+\tfrac{\lambda}{2}\right) \left( A_1-A_2^+ \right) + \lambda \left( B_2^+ - B_1 \right) \right) e^{-\tfrac{\lambda}{2}}  , \\
		g_4(\lambda) &= C^+ e^{\tfrac{\lambda}{2}} + \lambda \left( \left(1+\tfrac{\lambda}{2}\right) \left(A_3-A_2^-\right) + \lambda \left(B_3 - B_2^-\right) \right) e^{-\tfrac{\lambda}{2}}  ,
	\end{align}
\end{subequations}
wherein~$C^\pm = \lambda \left( \left(1-\lambda/2\right) A_2^\pm \mp \lambda B_2^\pm \right)$.

Inserting Eqs.~\eqref{X2} and \eqref{Q}, Eqs.~\eqref{InnerProblem} through~\eqref{g1234} form a system of four dual integral equations~\citep{tricomi85} for the unknown wavenumber-dependent coefficients regrouped in $\vect{X}_{13}$.
A solution of such types of dual integral equations with Bessel kernels can be obtained by the methods prescribed by \cite{sneddon60, sneddon66} and \cite{copson61}.
A similar procedure has recently been employed by some of us to address the axisymmetric flow induced by a Stokeslet near a circular elastic membrane~\citep{daddi19jpsj}, and the asymmetric flow field near a finite-sized rigid disk~\citep{daddi-compo20}.
Once~$\vect{X}_{13}$ is determined from solving the dual integral equations derived above, the remaining wavenumber-dependent coefficients expressed by~$\vect{X}_{2}$ follow forthwith from Eqs.~\eqref{X2} and~\eqref{Q}.

The core idea of our solution approach consists of expressing the solution of Eqs.~\eqref{OuterProblem} as definite integrals of the forms
\begin{subequations}\label{giDef}
	\begin{align}
		g_i(\lambda) &= 2\lambda^{\frac{1}{2}} \int_0^R f_i(t) J_{\frac{3}{2}} (\lambda t) \, \Intd t \, , \quad i \in \{1,3\} \, , \label{g_1_3} \\
		g_i(\lambda) &= 2\lambda^{\frac{1}{2}} \int_0^R f_i(t) J_{\frac{1}{2}} (\lambda t) \, \Intd t \, , \quad i \in \{2,4\} \, , \label{g_2_4}
	\end{align}
\end{subequations}
where~$f_i: [0,R] \to \mathbb{R}$, for $i \in \{1,2,3,4\}$, are unknown functions to be determined. 
Accordingly, the integral equations in the outer domain boundaries are automatically satisfied upon making use of the following identity, which holds for any positive integer~$p$~\citep{abramowitz72}, 
\begin{equation}
	\int_0^\infty \lambda^{\frac{1}{2}} J_p(\lambda r) J_{p+\frac{1}{2}} (\lambda t) \, \Intd \lambda = 0 \qquad (0<t<r) \, .
\end{equation}

By solving Eqs.~\eqref{g1234} for the coefficients~$A_1$, $B_1$, $A_3$, and~$B_3$ upon making use of Eqs.~\eqref{X2} and~\eqref{Q}, Eqs.~\eqref{InnerProblem} can be rewritten as
\begin{subequations}\label{InnerProblem2}
	\begin{align}
		 \int_{0}^{\infty} \left(2\lambda\right)^{-1} \left( g_{1}(\lambda) + \left( \lambda - 1 \right) e^{-\lambda} g_{3}(\lambda) + \lambda e^{-\lambda} g_{4}(\lambda) \right) J_{1}(\lambda r) \, \Intd \lambda &= \psi_{1}^{+}(r) \, , \\
		\int_{0}^{\infty} \left(2\lambda\right)^{-1} \left( \left( \lambda - 1 \right) e^{-\lambda} g_{1}(\lambda) + \lambda e^{-\lambda} g_{2}(\lambda) + g_{3}(\lambda) \right) J_{1}(\lambda r) \, \Intd \lambda &= \psi_{1}^{-}(r) \, , \\
		\int_{0}^{\infty} \left(2\lambda\right)^{-1} \left( g_{2}(\lambda) + \lambda e^{-\lambda} g_{3}(\lambda) + \left( \lambda + 1 \right) e^{-\lambda} g_{4}(\lambda) \right) J_{0}(\lambda r) \, \Intd \lambda &= \psi_{2}^{+}(r) \, , \\
		\int_{0}^{\infty} \left(2\lambda\right)^{-1} \left( \lambda e^{-\lambda} g_{1}(\lambda) + \left( \lambda + 1 \right) e^{-\lambda} g_{2}(\lambda) + g_{4}(\lambda) \right) J_{0}(\lambda r) \, \Intd \lambda &= \psi_{2}^{-}(r) \, .
	\end{align}
\end{subequations}

Next, by substituting Eqs.~\eqref{giDef} into Eqs.~\eqref{InnerProblem2}  and interchanging the order of the integrations with respect to the variables~$t$ and~$\lambda$, the equations associated with the inner problem can be expressed in the following final forms
\begin{subequations}\label{InnerProblem3}
	\begin{align}
		\int_0^R \left( L_5(r,t)f_1(t) + L_4(r,t)f_3(t) + L_1(r,t)f_4(t) \right) \Intd t &= \psi_1^+ (r) \, , \label{Eq_Psi1P} \\
		\int_0^R \left( L_4(r,t) f_1(t) + L_1(r,t)f_2(t) + L_5(r,t) f_3(t) \right) \Intd t &= \psi_1^-(r) \, ,\\
		\int_0^R \left( L_6(r,t) f_2(t) + L_3(r,t) f_3(t) + L_2(r,t)f_4(t) \right) \Intd t &= \psi_2^+(r) \, ,\\
		\int_0^R\left( L_3(r,t) f_1(t) + L_2(r,t)f_2(t) + L_6(r,t) f_4(t) \right) \Intd t &= \psi_2^-(r) \, , \label{Eq_Psi2N}
	\end{align}
\end{subequations}
where the kernels~$L_i: [0,R]^2 \to \mathbb{R}$, for $i \in \{1,2,3,4\}$ are complex mathematical functions that are defined and provided in Appendix~\ref{Appendix}.

Equations~\eqref{InnerProblem3} form a system of four Fredholm integral equations of the first kind~\citep{smithies58, polyanin98} for the unknown functions~$f_i(t)$, $i \in \{1,2,3,4\}$.
Due to the complicated nature of the kernel functions, we recourse to numerical solutions.

\subsection{Numerical solution of the integral equations and comparison with FEM simulations}

We now summarise the main steps involved in the numerical computations of the flow field.
First, the integration over the intervals~$[0,R]$ in Eqs.~\eqref{InnerProblem3} are partitioned into~$N$ subintervals and each integral is approximated by the standard middle Riemann sum~\citep{davis07}.
The four resulting equations are evaluated at~$N$ values of~$t_j$ that are uniformly distributed over the interval~$[0,R]$ such that $t_j = (j-1/2)(R/N)$, with $j = 1, \dots, N$.
Secondly, the discrete values of~$f_i (t_j)$, with $i \in \{1,2,3,4\}$ are obtained by solving the resulting linear system of~$4N$ equations. 
Thirdly, the four integrals in Eqs.~\eqref{giDef} are converted into well-behaved definite integrals over~$[0,\pi/2]$ by using the change of variable~$\lambda = \tan u$ and thus $\Intd \lambda = \Intd u / \cos^2 u$.
Thereupon, the resulting integrals are also approximated by the middle Riemann sum, and the wavenumber-dependent functions~$g_i(\lambda_k = \tan u_k)$, $k = 1, \dots, M$, are evaluated at discrete values of~$u_k$ such that $u_k = (k-1/2)(\pi/ 2 )/M$.
Fourthly, the values of $\vect{X}_2$ at each discrete point~$\lambda_k$ are readily obtained by inverting the linear system of four equations given by Eqs.~\eqref{g1234}.
In addition, it follows from Eq.~\eqref{X2} that $\vect{X}_{13} = \vect{Q}^{-1} \cdot \vect{X}_2$.
Finally, the image flow fields are obtained from Eqs.~\eqref{ExpressionsVelocityPressure} by approximating, again, the integrals by the middle Riemann sum.

\begin{figure}
	\centering
	\includegraphics[scale=1]{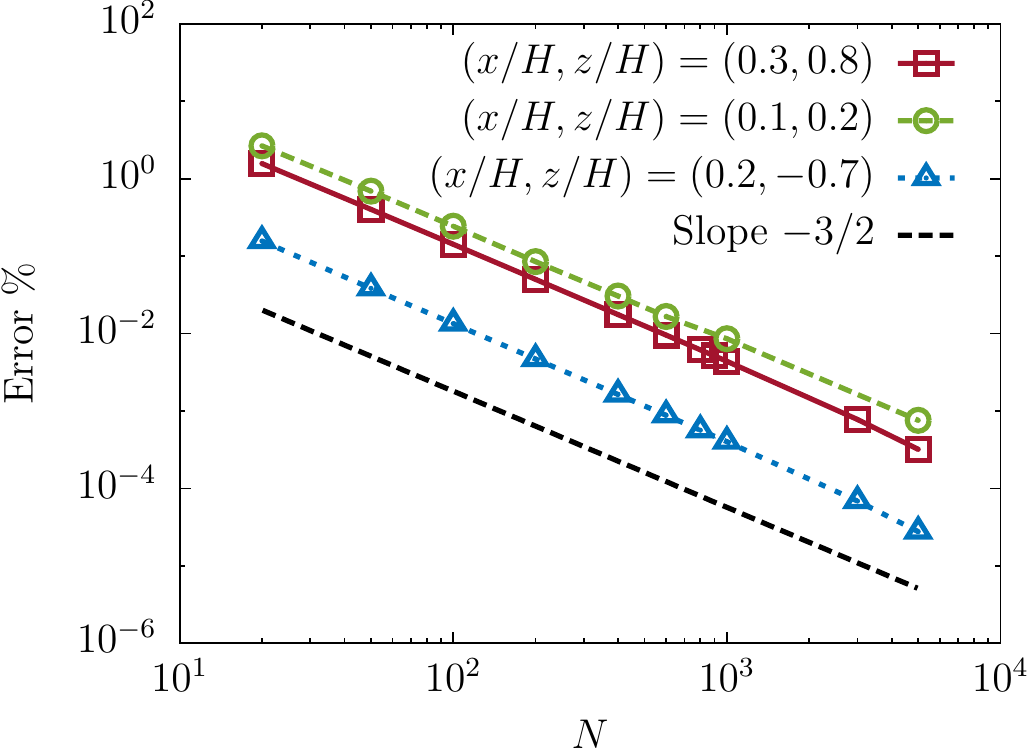}
	\caption{
	(Colour online) Log-log plot of the relative discretisation error occurring in the computation of the amplitude of the image velocity field versus the number of discretisation points, evaluated at various positions within the fluid domain. 
	Here, we set $R=H$, $h/H=0.3$, and $M = 10N$. 
	The errors are estimated relative to the corresponding values computed using a finer grid spacing with~$N=15000$ and $M=150000$.  
	}
	\label{Erro-Plot}
\end{figure}

Even though the approach employed here may seem cumbersome at a first glance, it has the advantage of being amenable to straightforward  implementation. 
Unlike many direct numerical simulation techniques which generally require discretisation of the entire three-dimensional fluid domain, or of at least an effectively two-dimenional domain when the axial symmetry is exploited, the integral formulation presented in this work reduces the solution of the flow problem to a set of one-dimensional integrals.
Besides, the present semi-analytical approach might serve as a motivation for various theoretical investigations of related problems that could possibly pave the way towards real engineering applications.

In Fig.~\ref{Erro-Plot}, we present a log-log plot the variations of the discretisation error~\citep{roy10} associated with the numerical computation of the amplitude of the image velocity field versus the number of discrete points used in the numerical integration of Eqs.~\eqref{InnerProblem3} while keeping $M=10N$ in the discretisation of Eqs.~\eqref{giDef} and~\eqref{ExpressionsVelocityPressure}.
The error is estimated relative to the numerical solution on a finer gird size for $N=15000$ and $M=150000$ at three different points of the fluid domain.
We observe that the error decays approximately algebraically as~$N^{-3/2}$ over the whole range of considered values of~$N$ and lies well below $10^{-3}$~$\%$ for $N \ge 5000$.
We have checked that a similar behaviour is also found when varying the position of the Stokeslet or the evaluation point within the fluid domain.

\begin{figure*}
	\centering
	\includegraphics[scale=1]{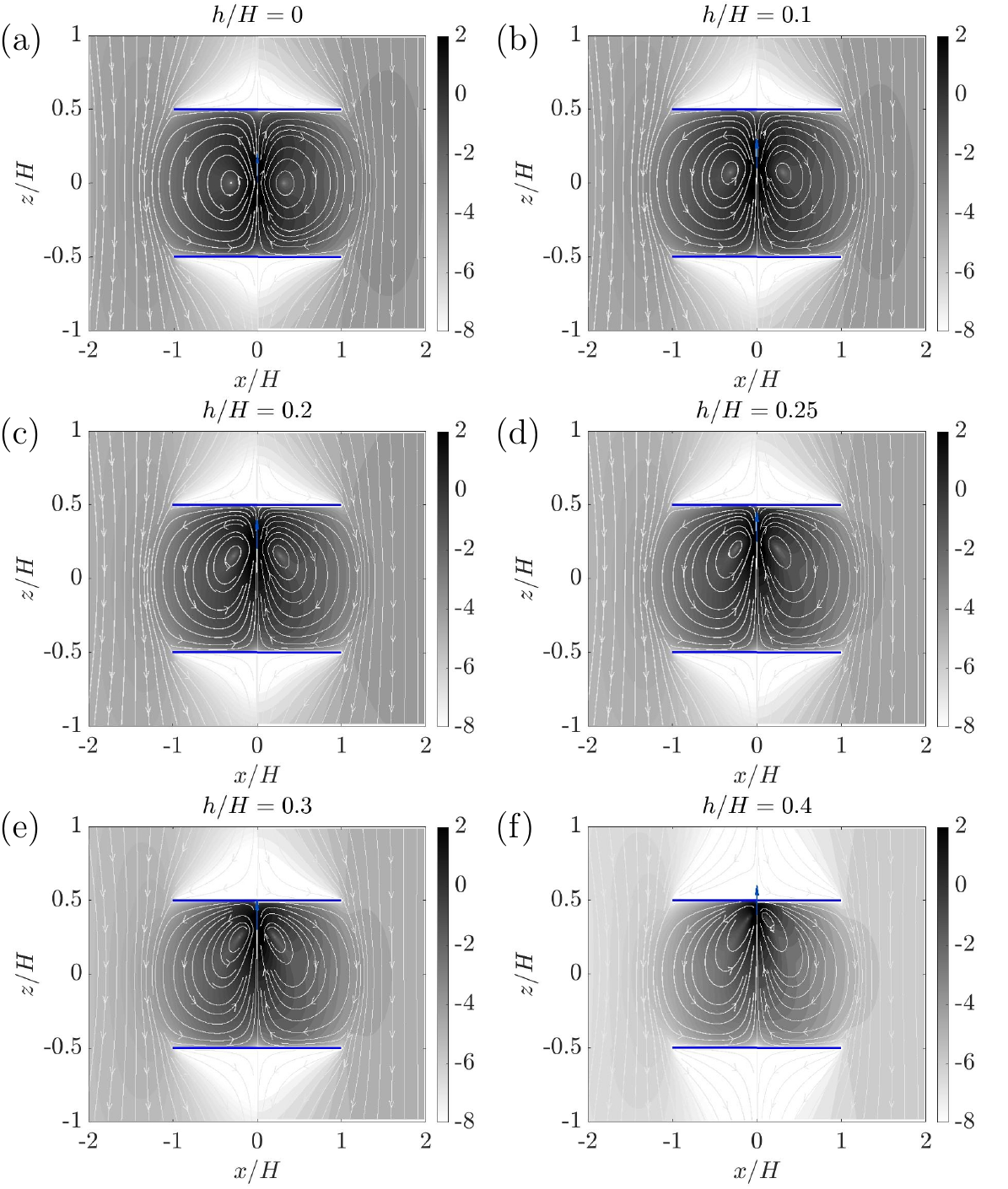}
	\caption{Streamlines and contour plots of the flow field induced by a point-force singularity acting inside two coaxially positioned disks of no-slip surfaces and of rescaled unit radius for various values of the vertical distance~$h/H$.
	In each panel, the flow velocity field obtained using the present semi-analytical approach is displayed in the left domain corresponding to~$x \le 0$, while the solution obtained using FEM simulations is presented in the right domain corresponding to~$x \ge 0$ for the same set of parameters.
	Here, we have defined the scaled flow velocity as $\vect{V} = \vect{v} / (F/(8\pi\eta)) $.}
	\label{Streamlines}
\end{figure*}

To validate our semi-analytical solution, we perform direct numerical simulations for the same geometry as well. 
We use a piecewise quadratic finite-element discretisation of the Stokes problem stated by Eqs.~\eqref{StokesGleischungenCylind} in cylindrical coordinates.
Since such an equal-order discretisation does not satisfy the inf-sup condition, we add stabilisation terms of local projection type~\citep{BeckerBraack2001}. 
The numerical domain is artificially limited to $(0,R)\times(-Z,Z)$ with $R,Z\in\mathbb{R}$ being sufficiently large numbers so as to avoid spurious feedback to the region of interest close to the plates.
In addition, the Dirac delta function forcing the flow is represented exactly in the variational formulation by means of
\begin{equation}
  \int_0^R\int_{-Z}^Z r \delta (\vect{r}-\vect{r}_0)\phi_z(\vect{r})\,\Intd r\,\Intd z \, 
  = \phi_z(\vect{r}_0),
\end{equation}
where $\phi_z$ is the test function corresponding to the vertical direction. Numerically, the singularity calls for very fine mesh resolution close to $\vect{r}_0$ and in proximity to the coaxially positioned plates, which we accomplish by local mesh adaptivity~\citep{BraackRichter2006d}. Further details on the discretisation method and the solution of the resulting linear systems of equations can be found in~\cite{Richter2017}.

In Fig.~\ref{Streamlines}, we represent the graphs of the resulting streamlines as well as contour plots of the total velocity field resulting from a Stokeslet singularity axisymmetrically acting at various positions along the axis of two coaxially disposed disks of unit radius.
Here, we set the numbers of discrete points to $N=15000$ and $M=150000$ in our numerical evaluation of the analytical description. 
The magnitude of the scaled velocity field is shown on a logarithmic scale in order to better appreciate the difference in magnitude between the different fluid regions.
In each panel, we depict on the left-hand side the results obtained via our semi-analytical approach derived in the present work. On the right-hand side in each panel, we include the corresponding flow fields determined via the FEM simulations.
Good agreement between the two solution procedures is obtained over the whole fluid domain, demonstrating the robustness and applicability of our semi-analytical approach.  
Most noticeably, we observe the existence of adjacent counterrotating eddies, the axis of rotation of which is directed along the azimuthal direction. 
Accordingly, the resulting flow field in the inner region consists of toroidal eddies on account of the axisymmetric nature of the flow~\citep{moffatt64}.
In contrast to that, descending streamlines are obtained in the outer region. 
For infinitely large disks, analogous toroidal structures have previously been identified and proven to decay exponentially with distance from the singularity position~\citep{liron81}.
Moreover, we remark that the overall magnitude of the flow field becomes less important as the point force gets closer to a confining plate.
This behaviour is accompanied by a notable increase of the asymmetry of the counterrotating eddies.

Having derived the solution of the flow problem due to an axisymmetric Stokeslet acting near two finite-sized coaxially positioned disks, we next employ our formalism to recover the solution earlier obtained by \cite{liron76} for a Stokeslet acting between two parallel planar walls of infinite extent along the transverse direction.

\subsection{Solution for $R\to\infty$}

For infinitely large disks, the integral equations~\eqref{InnerProblem2} in the inner domain become defined for the whole axis of positive real numbers.
Accordingly, the solution for the unknown functions $g_i(\lambda)$, for $i \in \{1,2,3,4\}$ can be obtained using inverse Hankel transforms.
By making use of the orthogonality property of Bessel functions~\citep{abramowitz72}
\begin{equation}
	\int_{0}^{\infty} r J_{\nu}(\lambda r) J_{\nu}(\lambda' r) \, \Intd r = \lambda^{-1} \DiracDelta{\lambda - \lambda'} \, ,
\end{equation}
we readily obtain
\begin{equation}
	\vect{H} \boldsymbol{\cdot} \vect{g} = \boldsymbol{\hat{\psi}} \, , \label{HgPsi}
\end{equation}
where we have defined the unknown vector $\vect{g} = \left(g_1, g_2, g_3, g_4\right)^\mathrm{T}$, the wavenumber-dependent matrix
\begin{equation} \label{H}
	\vect{H} = 
	\begin{pmatrix}
		e^\lambda & 0 & \lambda -1 & \lambda \\
		\lambda -1 & \lambda & e^\lambda & 0 \\
		0 & e^\lambda & \lambda & \lambda + 1 \\
		\lambda & \lambda + 1 & 0 & e^\lambda
	\end{pmatrix} \, , 
\end{equation}
and where~$\boldsymbol{\hat{\psi}} = \left(\hat{\psi}_{1}^{+}, \hat{\psi}_{1}^{-}, \hat{\psi}_{2}^{+}, \hat{\psi}_{2}^{-}\right)^\mathrm{T}$ gathers the inverse Hankel transforms of the previously introduced auxiliary functions defined by Eqs.~\eqref{psi}.
Specifically,
\begin{subequations}
	\begin{align}
		\hat{\psi}_{1}^{\pm}(\lambda) = \int_{0}^{\infty} r \psi_{1}^{\pm}(r) J_{1}(\lambda r) \, \Intd r &= \left( \tfrac{1}{2} \pm h \right) e^{-\lambda \left(\frac{1}{2} \pm h \right) } \, , \\
		\hat{\psi}_{2}^{\pm}(\lambda) = \int_{0}^{\infty} r \psi_{2}^{\pm}(r) J_{0}(\lambda r) \, \Intd r &= \left( \tfrac{1}{\lambda} + \tfrac{1}{2} \pm h \right) e^{- \lambda \left(\frac{1}{2} \pm h \right)} \, ,
	\end{align}
\end{subequations}
for $|h| < 1/2$. 
Solving the linear system of equations given by Eqs.~\eqref{HgPsi} and~\eqref{H} for the unknown vector function~$\vect{g}$ upon making use of Eqs.~\eqref{X2}, \eqref{Q}, and~\eqref{g1234} leads to 
\begin{equation}
	\vect{X}_{13} = \left( e^{-\lambda h}, -h e^{-\lambda h}, e^{\lambda h}, -h e^{\lambda h} \right)^\mathrm{T} \, .
\end{equation}

Accordingly, the total velocity and pressure fields in the lower and upper regions vanish in the limit $R\to\infty$.
The corresponding solution in the intermediate fluid domain can readily be obtained by invoking Eqs.~\eqref{X2} and~\eqref{Q}.

\section{Hydrodynamic mobility}
\label{sec:mobilitaet}

Our calculation of the flow field presented in the previous section can be employed in order to probe the effect of the two hard disks on the hydrodynamic drag acting on an enclosed point-like particle axially moving along the coaxially positioned axis.
This effect is commonly quantified by the hydrodynamic mobility function, which relates the velocity of a particle to the net force exerted on its surface~\citep{leal80, swan07,daddi16c,daddi17b,daddi2018brownian, driscoll19}.
In a bulk Newtonian fluid of constant dynamic viscosity~$\eta$, the mobility function~$\mu$ of a spherical particle of radius~$a$ is given by the familiar Stokes law, which states that in this case the mobility is $\mu_0 = 1/(6\pi\eta a)$~\citep{stokes51}.
In the presence of the confining disks, the leading-order correction to the particle mobility for an axisymmetric motion along the axis is obtained by evaluating the image flow field at the particle position as
\begin{equation}
	\Delta\mu = F^{-1}\lim_{(r,z) \to (0, h)} {v_z^*}_2 (r,z) \, . 
\end{equation}
Evaluating the limit in the latter equation and scaling by the bulk mobility, the scaled correction to the particle mobility is obtained as
\begin{equation}
	\frac{\Delta\mu}{\mu_0} = -k a \, , \label{k}
\end{equation}
where 
\begin{equation}
	k = \frac{3}{4} \int_0^\infty \Big(
	\big( \left(1-\lambda h\right)A_2^+ - \lambda B_2^+ \big) e^{\lambda h}
	+ \big( \left(1+\lambda h\right)A_2^- + \lambda B_2^- \big) e^{-\lambda h} \Big) \Intd \lambda \, \label{CorrectionFactor}
\end{equation}
is a positive dimensionless number commonly denominated as the correction factor of the Stokes steady mobility~\citep{happel12}.
Unfortunately, an analytical evaluation of this infinite integral is not auspicious.
Therefore, we recourse to a numerical evaluation.

For infinitely large disks, \textit{i.e.}, as $R\to\infty$, the correction factor~$k$ in Eq.~\eqref{k} can conveniently be cast into the simple integral form
\begin{equation}
	k_{\infty} = \frac{3}{8} \int_0^\infty W(\lambda) \left( \sinh^2\lambda - \lambda^2 \right)^{-1} \, \Intd \lambda \, , 
	\label{k-R-Inf}
\end{equation}
where we have defined the wavenumber-dependent function
\begin{equation}
	W(\lambda) = \Gamma_+ + \Gamma_- + \gamma_+ + \gamma_- 
	+ e^{-2\lambda} - \beta_+\beta_- \lambda^3 - 2\lambda^2 - 2\lambda - 1 \, ,
\end{equation}
with
\begin{equation}
	\beta_\pm = 1 \pm 2h \, , \quad
	\Gamma_\pm = \left( 1 + \tfrac{1}{2} \, \lambda^2 \beta_\pm^2 \right) \sinh \left( \lambda \beta_\mp \right) \, , \quad
	\gamma_\pm = \lambda \beta_\pm \cosh \left( \lambda \beta_\mp \right) \, .
\end{equation}
This result is found to be in full agreement with the expression obtained by \cite{swan10}, who used a two-dimensional Fourier transform technique.

\begin{figure}
	\begin{center}
		\includegraphics[scale=0.9]{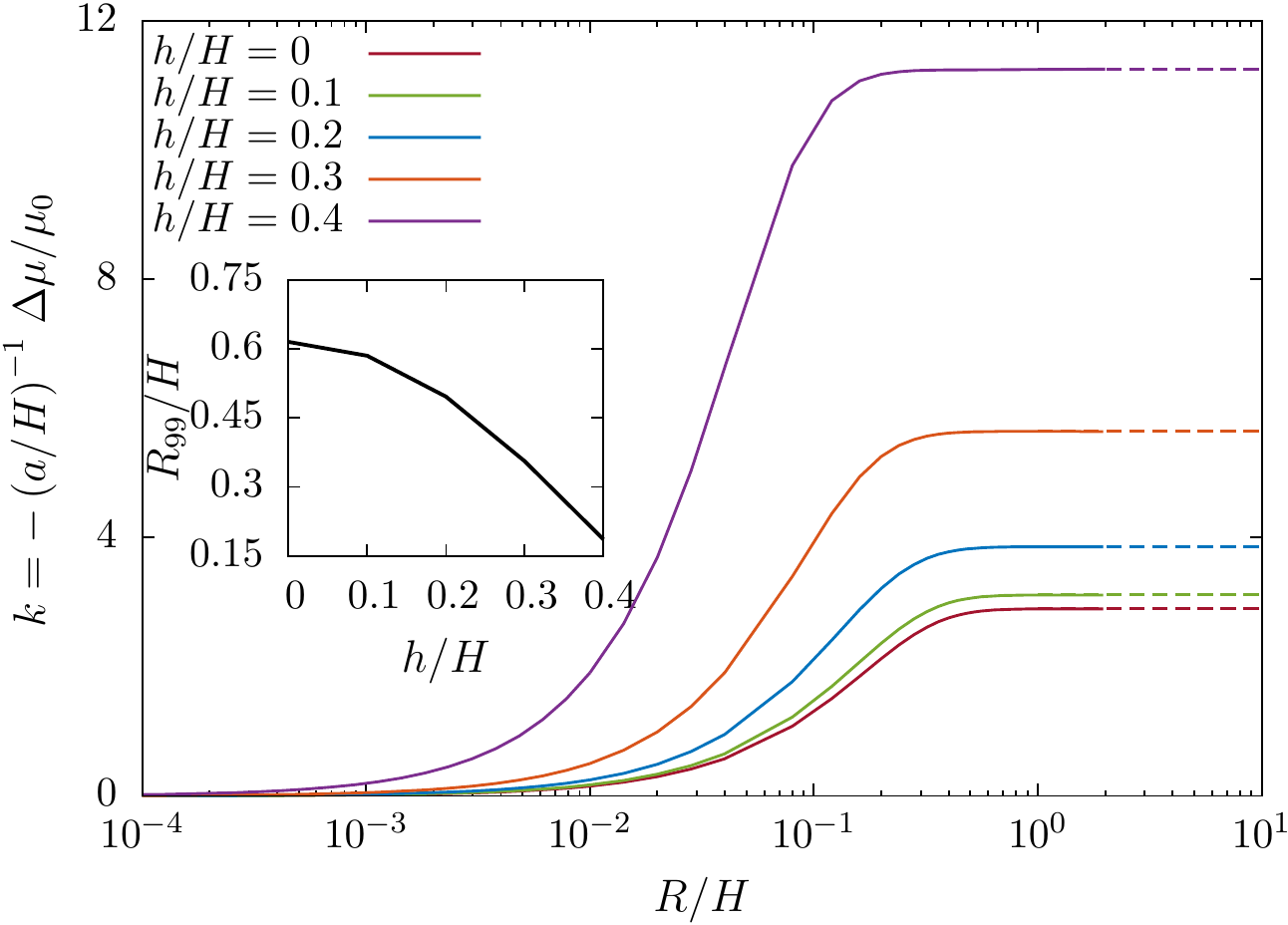}
		\caption{(Colour online) Variations of the correction factor of the hydrodynamic mobility as defined by Eq.~\eqref{CorrectionFactor} versus~$R/H$ for various values of~$h/H$. 
		Horizontal dashed lines correspond to the correction factor near two infinitely large disks as given by Eq.~\eqref{k-R-Inf}.
		Inset: Evolution of $R_{99}/H$ versus~$h/H$, where $R_{99}$ is defined such that $k \left(R_{99}/H\right) = 0.99k_\infty$, for which the correction factor near infinitely large disks is almost recovered.
		}
		\label{Mobi-Plot}
	\end{center}
\end{figure}

In Fig.~\ref{Mobi-Plot}, we present a linear-logarithmic plot of the correction factor of the mobility function versus the radius of the disks for various values of the singularity position.
Results are obtained by integrating Eq.~\eqref{CorrectionFactor} numerically.
We observe that the curves follow a sigmoid-logistic-like phenomenology, implying that the correction factor increases significantly in the range of small radii before it reaches a saturation value.
The latter corresponds to the correction factor predicted near two infinitely large disks given by Eq.~\eqref{k-R-Inf}.

Next, in order to quantify the effect of finite disk size on the correction to the hydrodynamics mobility, we customarily define the radius~$R_{99}$ for which the mobility near infinitely large disks is essentially reached, such that $k(R_{99}) = 0.99 k_\infty$. 
In the inset of Fig.~\ref{Mobi-Plot}, we display the variations of $R_{99}$ versus $h$ based on the data presented in the main plot.
We observe that $R_{99}$ reaches a maximum value of about 0.62 at the mid-plane of the channel before it monotonically decreases with $h$.
This observation suggests that, to a good approximation, the mobility near two infinitely large disks can adequately be used to estimate the mobility at arbitrary position along the axis provided that the radius-to-channel-height ratio is above 0.62.
Hence, accounting for the finite-size effect here becomes crucial only for values below this threshold.

Finally, we comment on the applicability of the often-used approximation originally suggested by \cite{oseen28} to predict the particle mobility between two boundaries by superimposing separately the leading-order effects of each boundary.
Accordingly,
\begin{equation}
	\frac{\Delta\mu_\mathrm{Sup}}{\mu_0} =  -k_\mathrm{Sup} a \, , \qquad
	k_\mathrm{Sup} = -a^{-1} \left( \left. \frac{\Delta\mu_\mathrm{Disk}}{\mu_0} \right|_{b=\frac{1}{2}-h} 
		+ \left. \frac{\Delta\mu_\mathrm{Disk}}{\mu_0} \right|_{b=\frac{1}{2}+h} \right) \, , 
	\label{k_Supo}
\end{equation}
where the leading-order correction to the mobility function for axisymmetric motion normal to one rigid circular disk has previously been obtained by \cite{kim83} and is expressed by
\begin{equation}
	\frac{\Delta\mu_\mathrm{Disk}}{\mu_0} = -\frac{3}{4\pi} 
	\left( \frac{3+5\xi^2}{\left(1+\xi^2\right)^2} + \frac{3}{\xi} \arctan \left( \frac{1}{\xi} \right) \right) \frac{a}{R} \, , 
\end{equation}
wherein~$\xi = b/R$ is a dimensionless parameter with $b$ denoting the distance between the particle and the centre of the disk.
This solution was obtained by formulating the flow problem in terms of a mixed-boundary-value problem and solving the resulting dual integral equations using an approach analogous to that employed in the present work. 
Notably, we recover for $\xi \to 0$ the familiar correction to the hydrodynamic mobility near an infinitely extended plane solid wall of no-slip boundary condition at its surface, namely $\Delta\mu_\mathrm{Disk}/\mu_0 = -9a/(8b)$, as originally obtained by Lorentz using the reciprocal theorem more than a century ago~\citep{lorentz07, lee79}.

\begin{figure}
	\begin{center}
		\includegraphics[scale=0.9]{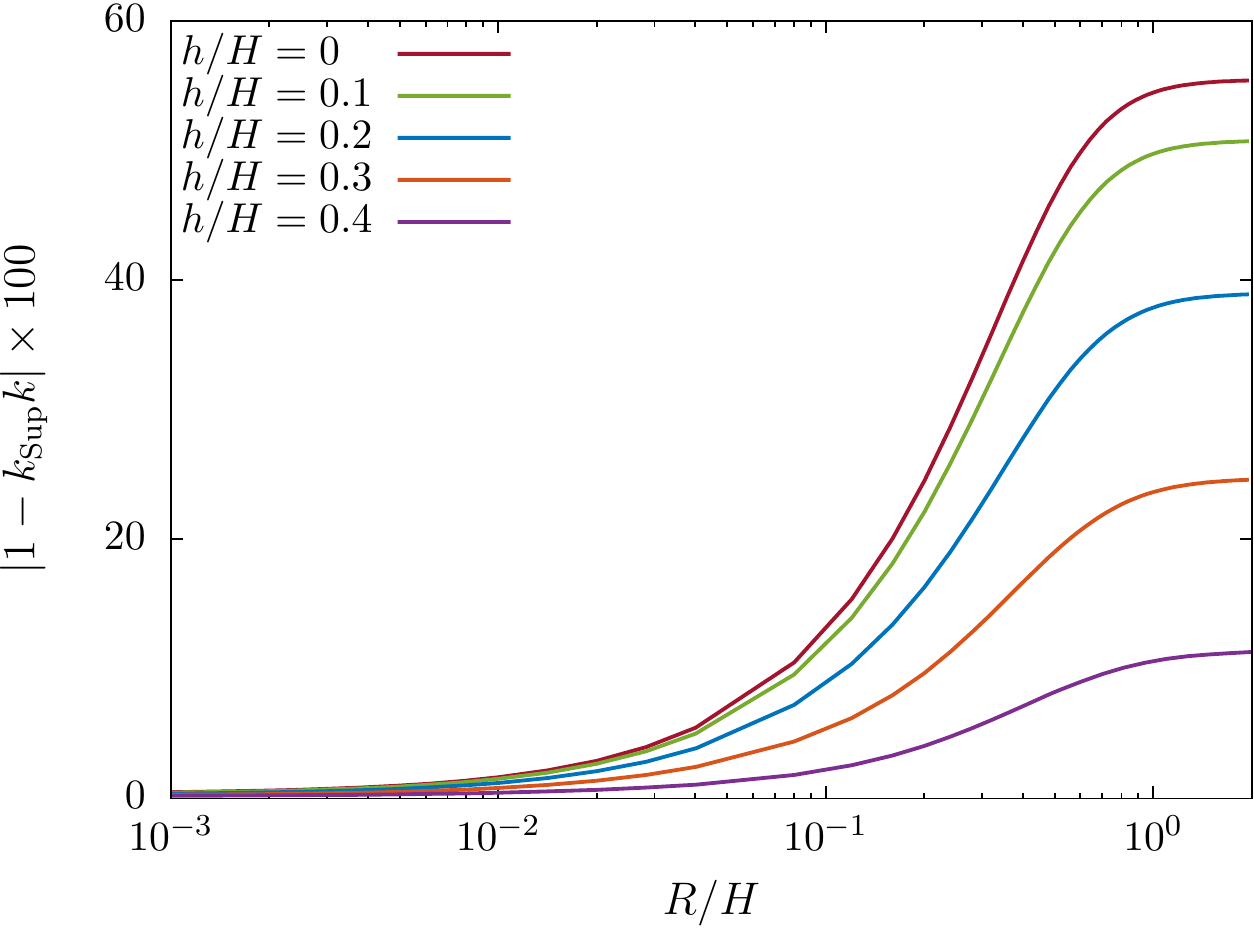}
		\caption{(Colour online)
		Percentage relative error between the correction factor of the Stokes steady mobility as obtained from the superposition approximation given by Eq.~\eqref{k_Supo} and the exact expression given by Eq.~\eqref{CorrectionFactor}.
		}
		\label{Supo-Err}
	\end{center}
\end{figure}

We now assess the accuracy of the superposition approximation stated by Eq.~\eqref{k_Supo} by direct comparison with the exact prediction given by Eq.~\eqref{CorrectionFactor}.
In Fig.~\ref{Supo-Err}, we plot the variations of the percentage relative error between the correction factors $k_\mathrm{Sup}$ and~$k$ versus the radius of the disks~$R$ for various values of the particle position~$h$.
In the range of small values of~$R$, the relative error amounts to small values, typically smaller than~10\% for $R < 0.1$.
Upon increasing~$R$, the relative error gradually increases in a logistic-like manner, before it saturates on a plateau value as $R$ gets larger.
The maximum error is obtained for the particle located on the mid-plane between the two disks for $h=0$ and is found to be of about~55\% in the limit of infinite disk radius.
Therefore, the superposition approximation cannot be applied properly in this case.
Nonetheless, as the particle position gets closer to either disk, the maximum error notably decreases to amount to only about~12\% for $h = 0.4$.
Consequently, the superposition approximation can frequently be utilised in this range of values to predict the hydrodynamic mobility for axisymmetric motion along the axis of the disks.

\section{Conclusions}
\label{sec:conclusions}

To summarise, we have examined the axisymmetric Stokes flow resulting from a Stokeslet singularity acting on the axis of two coaxially positioned circular disks of equal radius. 
We have formulated the solution for the viscous incompressible flow field as a mixed-boundary-value problem, which we have then reduced into a system of dual integral equations for four unknown wavenumber-dependent functions.
Most importantly, we have shown the existence of viscous toroidal eddies in the fluid region bounded by the two plates.
In the limit of infinitely large disks, we have successfully recovered the classic solution by \cite{liron76} for a Stokeslet acting normal to two parallel planar walls.

Additionally, we have provided an integral expression of the hydrodynamic mobility function quantifying the effect of the confining plates on the motion of a point-like particle moving along the axis of the coaxially positioned disks.
Furthermore, we have demonstrated that accounting for the finite-size effect of the disks becomes essential only below a threshold value of the radius-to-channel height.
Beyond this value, the mobility near two infinitely large disks can appropriately be employed.
Finally, we have tested the validity and robustness of Oseen's approximation that postulates that the particle mobility between two boundaries could approximately be predicted by superimposing the contributions from each boundary independently.
We have found that this simplistic approximation works quite well as the particle gets closer to either boundary but severely breaks down when the particle is located in the mid-plane between the two disks.

The analytical approach in the present article is based on the assumption of flow axisymmetry.
The Stokes flow induced by a Stokeslet directed along an arbitrary direction in the presence of two coaxially positioned disks would be worth being investigated in a future study.
We conjecture that this solution might be obtained by making use of the Green and Neumann functions supplemented by the edge function following the approach by \cite{miyazaki84}.
This solution can then be employed to evaluate the translational and rotational mobility functions of particles located at arbitrary positions between the two disks.
Alternatively, the problem can possibly be approached differently by means of multipole expansion methods involving the expression of the relevant hydrodynamic fields using oblate spheroidal coordinates~\citep{lee80}.
This approach has widely been employed in the context of micromechanics of heterogeneous composite materials and fracture analysis~\citep{kushch00, kushch13}.
In principle, our calculations can be extended to account for higher-order correction factors in the aspect ratio between the radius of the disks and the distance between the particle and the bounding plates~\citep{swan10}, but this would require a very challenging effort.

For applications requiring the precise manipulation of single molecules at the nanoscale level, the no-slip boundary condition may need to be lifted.
In this context, the effect of partial slip at the surfaces of the disks is commonly characterised by assuming that the velocity components of the fluid tangent to the surfaces of the disks is proportional to the rate of strain at the surfaces~\citep{lauga05,lasne2008velocity}.
This is an interesting aspect that could be included in our formalism and represents a worthwhile extension of the problem for future studies.
We hope that our study will prove useful to researchers as well as practitioners working on particulate flow problems involving finitely sized boundaries, and pave the way toward better design and control of various processes in micro- and nanofluidic systems.

\begin{acknowledgments}
	A.D.M.I., H.L., and A.M.M. gratefully acknowledge support from the DFG (Deutsche Forschungsgemeinschaft) through the projects DA~2107/1-1, LO~418/16-3, and ME~3571/2-2.
	
	Declaration of Interests. The authors report no conflict of interest.
\end{acknowledgments}

\appendix

\section{Analytical expressions for the kernel functions}
\label{Appendix}

In this Appendix, we provide technical details regarding the analytical derivation of the kernel functions appearing in the system of Fredholm integral equations of the first kind given by Eqs.~\eqref{InnerProblem3} of the main body of the article.

The kernel functions can be expressed as infinite integrals over the wavenumber~$\lambda$ as
\begin{subequations}\label{I123456}
	\begin{align}
		L_1 (r,t) &= \int_0^\infty \lambda^{\frac{1}{2}}
		e^{-\lambda} J_1 (\lambda r) J_{\frac{1}{2}} (\lambda t) \, \mathrm{d}\lambda \, , \label{I1} \\
		L_2 (r,t) &= \int_0^\infty \left( \lambda^{\frac{1}{2}}+\lambda^{-\frac{1}{2}} \right)
		e^{-\lambda} J_0 (\lambda r) J_{\frac{1}{2}} (\lambda t) \, \mathrm{d}\lambda \, , \label{I2} \\
		L_3 (r,t) &= \int_0^\infty \lambda^{\frac{1}{2}}
		e^{-\lambda} J_0 (\lambda r) J_{\frac{3}{2}} (\lambda t) \, \mathrm{d} \lambda\, , \label{I3} \\
		L_4 (r,t) &= \int_0^\infty \left( \lambda^{\frac{1}{2}}-\lambda^{-\frac{1}{2}} \right)
		e^{-\lambda} J_1 (\lambda r) J_{\frac{3}{2}} (\lambda t) \, \mathrm{d}\lambda \, , \label{I4} \\
		L_5(r,t) &= \int_0^\infty \lambda^{-\frac{1}{2}} J_1(\lambda r) J_{\frac{3}{2}} (\lambda t) \, \Intd \lambda \, , \label{I5} \\
		L_6(r,t) &= \int_0^\infty \lambda^{-\frac{1}{2}} J_0(\lambda r) J_{\frac{1}{2}} (\lambda t) \, \Intd \lambda \, , \label{I6}
	\end{align}
\end{subequations}
where $(r,t) \in [0,R]^2$.
It can be shown that the first four integrals can conveniently be expressed in closed mathematical forms as
\begin{subequations}\label{I1234}
	\begin{align}
		L_1(r,t) &= \left( \tfrac{2}{\pi t} \right)^\frac{1}{2} \tfrac{1}{r} \operatorname{Im}
		\left( \xi_+ \delta_+ \right) \, , \label{I1Res} \\
		L_2(r,t) &= \left( \tfrac{2}{\pi t} \right)^\frac{1}{2} 
		\left( \operatorname{Re} \left( \Lambda \right) + \operatorname{Im} \left( \delta_- \right) \right) \, , \label{I2Res} \\
		L_3(r,t) &= \left( \tfrac{2}{\pi t} \right)^\frac{1}{2} 
		\operatorname{Re} \left( \Lambda t^{-1} - \delta_- \right) \, , \label{I3Res} \\
		L_4(r,t) &= \left( \tfrac{2}{\pi t} \right)^\frac{1}{2}
		\left(\operatorname{Re}\left( \chi_1 \right) 
		+ \operatorname{Im} \left( \chi_2 \right) \right) \, , \label{I4Res} 
	\end{align}
\end{subequations}
where we have defined the abbreviations
\begin{subequations}
	\begin{align}
		\xi_\pm &= 1 \pm it \, , \quad
		\delta_\pm = \left( r^2 + \xi_\pm^2 \right)^{-\frac{1}{2}} \,, \quad 
		\Lambda = \arcsin \left( \frac{t+i}{r} \right) \, , \quad
		\sigma = \frac{r}{\xi_- + \delta_-^{-1}} \, , \label{greatCoeffs1} \\
		\alpha &= \frac{t}{r} \, , \quad
		\chi_1 = \delta_- \left( \tfrac{r}{2} \left(1+\sigma^2\right) + \tfrac{\xi_-}{r} \right) - \tfrac{\Lambda}{2\alpha} \,, \quad 
		\chi_2 = \tfrac{1}{rt\delta_-} + \tfrac{\delta_-}{8\alpha} \left( \xi_- - r\sigma^3 \right) \, . \label{greatCoeffs2}
	\end{align}
\end{subequations}

In addition, the integrals~$L_5$ and~$L_6$ have analytical forms and can be calculated directly form standard integration tables or software algebra systems such as Mathematica~\citep{wolfram99} as
\begin{subequations}
	\begin{align}
		L_5(r,t) &= \tfrac{1}{2} \left( \tfrac{\pi}{2t} \right)^\frac{1}{2} \alpha^{-1} H(t-r)
		+ \left( \tfrac{1}{2\pi t} \right)^\frac{1}{2} \left( \alpha^{-1} \arcsin \left(\alpha\right) - \left( 1-\alpha^2 \right)^\frac{1}{2} \right) H (r-t) \, , \\
		L_6(r,t) &= \left( \tfrac{\pi}{2t} \right)^\frac{1}{2} H(t-r)
		+ \left( \tfrac{2}{\pi t} \right)^\frac{1}{2} \arcsin (\alpha) H(r-t) \, ,
	\end{align}
\end{subequations}
where~$H(\cdot)$ denotes the Heaviside step function.


In the following, we will show how the integrals given by Eqs.~\eqref{I123456} can be evaluated analytically.
The core idea of our approach consists of expressing these integrals in the form of Laplace transforms of Bessel functions of the first kind~\citep{spiegel65, widder15},
\begin{equation}
	 \mathcal{L}\left\{ J_k(z)\right\} (p) = \left( 1+p^2 \right)^{-\halb} 
	 \left( p + \left(1+p^2\right)^\halb \right)^{-k} \, , \label{laplaceJ} 
\end{equation} 
and using the recurrence relation~\citep{abramowitz72}
\begin{equation}
	\frac{2k}{z} \, J_k(z) = J_{k-1}(z) + J_{k+1} (z) \, . \label{recurrence}
\end{equation}

In addition, we will employ the following identities providing closed-form expressions for the Bessel functions of the first kind of half-integer order in terms of the standard trigonometric functions,
\begin{subequations}\label{identity1UND2}
	\begin{align}
	 J_{\halb}(z) &= \left(\tfrac{2}{\pi z}\right)^\halb \sin(z) \, , \label{identity1} \\
	 J_{-\halb}(z) &= \left(\tfrac{2}{\pi z}\right)^\halb \cos(z) \, . \label{identity2}
	\end{align} 
\end{subequations}

\subsection*{Evaluation of the integral~$L_1$}

By making use of the identity given by Eq.~\eqref{identity1}, the integral~$L_1$ stated by Eq.~\eqref{I1} can be expressed as
\begin{equation}
 L_1 (r,t) = \left(\tfrac{2}{\pi t}\right)^\halb \int_0^\infty e^{-\lambda} J_1 (\lambda r)\sin{(\lambda t)} \, \mathrm{d}\lambda \, .
\end{equation}
Using the change of variable~$x=\lambda r$ and Euler's representation of the sine function, the latter integral can be expressed as
\begin{equation}
	L_1(r,t) = \left(\tfrac{2}{\pi t}\right)^\halb \tfrac{1}{r} \operatorname{Im} \left(\int_0^\infty e^{-\frac{x}{r}\left( 1-it \right)} J_1 (x) \, \mathrm{d}x \right) .
	\label{I1ResAppendix}
\end{equation}
This leads to Eq.~\eqref{I1Res} after making use of the Laplace transform given by Eq.~\eqref{laplaceJ} for $k=1$ and $p=(1-it)/r$.
We note that $\operatorname{Im}(z) = -\operatorname{Im}(\bar{z})$ for $z \in \mathbb{C}$, where $\bar{z}$ denotes the complex conjugate of~$z$.

\subsection*{Evaluation of the integral~$L_2$}

We next consider the integral defined by Eq.~\eqref{I2}, which can conveniently be decomposed into two parts as 
\begin{equation}
	L_2(r,t) = L_{2,1} (r,t) + L_{2,2} (r,t) \, , 
\end{equation}
where 
\begin{subequations}\label{I21I22}
	\begin{align}
		L_{2,1} (r,t) &= \left(\tfrac{2}{\pi t}\right)^{\halb} \int_0^\infty e^{-\lambda} J_0 (\lambda r)\sin{(\lambda t)} \, \mathrm{d}\lambda \, , \label{I21Int} \\
		L_{2,2} (r,t) &= \left(\tfrac{2}{\pi t}\right)^{\halb} \int_0^t \Intd u \int_0^\infty e^{-\lambda} J_0 (\lambda r)\cos{(\lambda u)} \, \mathrm{d}\lambda \, . 
	\end{align}
\end{subequations}
Here, we have made use of Eq.~\eqref{identity1} together with the integral representation
\begin{equation}
	\sin(\lambda t) = \lambda \int_0^t \cos(\lambda u) \, \Intd u \, . \label{sineIntRepres}
\end{equation}

Using Euler's relation together with Eq.~\eqref{laplaceJ} for $k=0$, Eqs.~\eqref{I21I22} can be evaluated as
\begin{subequations}
	\begin{align}
		L_{2,1} (r,t) &= \left(\tfrac{2}{\pi t}\right)^{\halb} \operatorname{Im}\left( \left(r^2+(1-it)^2\right)^{-\halb} \right) \, , \\
		L_{2,2} (r,t) &= \left(\tfrac{2}{\pi t}\right)^{\halb} \operatorname{Re} \left( \int_0^t \left(r^2+(1-iu)^2\right)^{-\halb} \, \mathrm{d}u \right) \, . \label{I22}
	\end{align}
\end{subequations}
The definite integral in Eq.~\eqref{I22} can be evaluated as
\begin{equation}
	L_{2,2} (r,t) = \left(\tfrac{2}{\pi t}\right)^{\halb} \operatorname{Re} \left( \arcsin \left( \frac{t+i}{r} \right) \right) \, . \label{I22ResAppendix}
\end{equation}
Equation~\eqref{I2Res} follows forthwith after collecting terms.

It is worth mentioning that, for a given complex number $z=x+iy$, the arcsine function is defined when $\pm x \notin (1,\infty)$ as~\citep{abramowitz72}
\begin{equation}
	\arcsin (z) = \arcsin (\alpha_-) + i \operatorname{sign}(y) \ln \left( \alpha_+ + \left( \alpha_+^2 - 1 \right)^{\halb} \right) \, , 
\end{equation}
where \begin{equation}
	\alpha_\pm = \tfrac{1}{2} \left((x+1)^2+y^2\right)^\halb
	 \pm \tfrac{1}{2} \left((x-1)^2+y^2\right)^\halb \, .
\end{equation}

\subsection*{Evaluation of the integral~$L_3$}

Analogously, the integral~$L_3$ defined by Eq.~\eqref{I3} can be decomposed into two parts as
\begin{equation}
	L_3(r,t) = L_{3,1}(r,t) - L_{3,2}(r,t) \, , 
\end{equation}
upon using the recurrence relation stated by Eq.~\eqref{recurrence} and setting~$k=1/2$ together with the identities given by Eqs.~\eqref{identity1UND2}.
Here, we have defined $L_{3,1}(r,t) = t^{-1} L_{2,2} (r,t)$ and 
\begin{equation}
	L_{3,2}(r,t) = \left( \tfrac{2}{\pi t} \right)^\halb
	\int_0^\infty e^{-\lambda} J_0(\lambda r) \cos(\lambda t) \, \Intd \lambda \, ,
\end{equation}
which can readily be evaluated as Eq.~\eqref{I21Int} but by taking this time the real part.
This leads to Eq.~\eqref{I3Res} upon collecting terms.

\subsection*{Evaluation of the integral~$L_4$}

Finally, upon using Eq.~\eqref{recurrence} for~$k=1/2$ and the identities given by Eqs.~\eqref{identity1UND2}, the integral~$L_4$ can be decomposed into four parts
\begin{equation}
L_4 (r,t) = L_{4,1} + L_{4,2} - \left( L_{4,3} + L_{4,4} \right) \, , 
\end{equation}
where we have defined
\begin{subequations}
	\begin{align}
		L_{4,1} (r,t) &= \left( \tfrac{2}{\pi t} \right)^\halb t^{-1} \int_0^\infty \lambda^{-1} e^{-\lambda} J_1 (\lambda r)\sin{(\lambda t)} \, \mathrm{d}\lambda \, , \\
		L_{4,2} (r,t) &= \left(\tfrac{2}{\pi t}\right)^\halb \int_0^\infty \lambda^{-1} e^{-\lambda} J_1 (\lambda r)\cos{(\lambda t)} \, \mathrm{d}\lambda \, , \\
		L_{4,3} (r,t) &= \left(\tfrac{2}{\pi t}\right)^\halb \int_0^\infty e^{-\lambda} J_1 (\lambda r)\cos{(\lambda t)} \, \mathrm{d}\lambda \, , \\
		L_{4,4} (r,t) &= \left( \tfrac{2}{\pi t} \right)^\halb t^{-1} \int_0^\infty \lambda^{-2} e^{-\lambda} J_1 (\lambda r)\sin{(\lambda t)} \, \mathrm{d}\lambda \, .
	\end{align}
\end{subequations}

In the following, we will make use when appropriate of the shorthand notations defined in Eq.~\eqref{greatCoeffs1}.
By using the integral representation of the sine function given by Eq.~\eqref{sineIntRepres}, the first integral can be expressed as
\begin{equation}
	L_{4,1} (r,t) = \left( \tfrac{2}{\pi t} \right)^\halb t^{-1} \int_0^t\,\mathrm{d} u\int_0^\infty e^{-\lambda} J_1 (\lambda r)\cos{(\lambda u)} \, \mathrm{d}\lambda \, .
\end{equation}
Similarly, the evaluation of the indefinite integral over~$\lambda$ can be performed using the Laplace transform of the Bessel function given by Eq.~\eqref{laplaceJ} to obtain
\begin{align}
	L_{4,1} (r,t) = \left( \tfrac{2}{\pi t} \right)^\halb (tr)^{-1} 
	\operatorname{Re} \left(
	\int_0^t \left(1 - (1-iu) \left(r^2+(1-iu)^2\right)^{-\halb} \right) \mathrm{d} u \right) .
\end{align}
The definite integral in the latter equation can then be evaluated and cast in the final simplified form
\begin{equation}
	L_{4,1} (r,t) = \left( \tfrac{2}{\pi t} \right)^\halb r^{-1} \left(1 + t^{-1} \operatorname{Im}\left( \delta_-^{-1}\right)\right) \, .
\end{equation}

Next, the evaluation of the second integral is straightforward after expressing the first-order Bessel function as a function of the zeroth and second order Bessel functions using the recurrence relation given by Eq.~\eqref{recurrence} for~$k=1$ to obtain
\begin{equation}
L_{4,2} (r,t) = r \left(2\pi t\right)^{-\halb} \int_0^\infty  e^{-\lambda} \left( J_0 (\lambda r) + J_2(\lambda r) \right) \cos{(\lambda t)} \, \mathrm{d}\lambda \, , 
\end{equation}
which can readily be evaluated as
\begin{equation}
	L_{4,2} (r,t) = r \left(2\pi t\right)^{-\halb} \operatorname{Re}\left(
	\delta_- \left( 1 + \sigma^2 \right) \right) \, .
\end{equation}

The third integral can be deduced from the calculation of $L_1 (r,t)$ given by Eq.~\eqref{I1ResAppendix}, this time by taking the real part to obtain
\begin{equation}
	L_{4,3}= \left(\tfrac{2}{\pi t}\right)^\halb r^{-1}\operatorname{Re}\left(1- \xi_- \delta_- \right) .
\end{equation} 

Lastly, the fourth integral can be decomposed into two parts as
\begin{equation}
	L_{4,4} (r,t) = L_{4,4,1} (r,t) + L_{4,4,2} (r,t) \, , 
\end{equation}
where $L_{4,4,1} (r,t) = (2\alpha)^{-1} L_{2,2} (r,t)$ and
\begin{equation}
		L_{4,4,2} (r,t) = (2\pi t)^{-\halb} \tfrac{r}{t}
		\int_0^\infty \lambda^{-1} e^{-\lambda} J_2 (\lambda r)\sin{(\lambda t)} \, \mathrm{d}\lambda \, .
\end{equation}
This integral can be handled using the recurrence formula given by Eq.~\eqref{recurrence} to obtain
\begin{equation}
L_{4,4,2} (r,t)= (2\pi t)^{-\halb} 
\tfrac{r^2}{4t}\int_0^\infty e^{-\lambda} \left(J_1 (\lambda r) + J_3(\lambda r) \right)\sin{(\lambda t)} \, \mathrm{d}\lambda \, .
\end{equation}
The latter integral can be calculated and cast in the final simplified form
\begin{equation}
	L_{4,4,2} (r,t) = (2\pi t)^{-\halb} \left(4\alpha\right)^{-1} \left(
	r\operatorname{Im}\left(\delta_- \sigma^3 \right)
	-\operatorname{Im}\left(\xi_- \delta_-\right) \right) \, .
\end{equation}
By collecting terms, Eq.~\eqref{I4Res} is readily obtained.


\begin{thebibliography}{121}
\expandafter\ifx\csname natexlab\endcsname\relax\def\natexlab#1{#1}\fi
\def\au#1{#1} \def\ed#1{#1} \def\yr#1{#1}\def\at#1{#1}\def\jt#1{\textit{#1}}
  \def\bt#1{#1}\def\bvol#1{\textbf{#1}} \def\vol#1{#1} \def\pg#1{#1}
  \def\publ#1{#1}\def\arxiv#1{#1}\def\org#1{#1}\def\st#1{\textit{#1}}

\bibitem[Abramowitz \& Stegun(1972)]{abramowitz72}
{\sc \au{Abramowitz, M.} \& \au{Stegun, I.~A.}} \yr{1972} {\em {Handbook of
  Mathematical Functions}\/}.  \publ{Dover, New York, U.S.A.}

\bibitem[Asta {\em et~al.\/}(2019)Asta, Palaia, Trizac, Levesque \&
  Rotenberg]{asta19}
{\sc \au{Asta, A.~J.}, \au{Palaia, I.}, \au{Trizac, E.}, \au{Levesque, M.} \&
  \au{Rotenberg, B.}} \yr{2019}  \at{Lattice {Boltzmann} electrokinetics
  simulation of nanocapacitors}.  \jt{J. Chem. Phys.}  \bvol{151}~(11),
  \pg{114104}.

\bibitem[Babel {\em et~al.\/}(2018)Babel, Eikerling \& L\"{o}wen]{babel18}
{\sc \au{Babel, S.}, \au{Eikerling, M.} \& \au{L\"{o}wen, H.}} \yr{2018}
  \at{Impedance resonance in narrow confinement}.  \jt{J. Phys. Chem. C}
  \bvol{122}~(38),  \pg{21724--21734}.

\bibitem[Balducci {\em et~al.\/}(2006)Balducci, Mao, Han \& Doyle]{balducci06}
{\sc \au{Balducci, A.}, \au{Mao, P.}, \au{Han, J.} \& \au{Doyle, P.~S.}}
  \yr{2006}  \at{{Double-stranded DNA diffusion in slitlike nanochannels}}.
  \jt{Macromolecules}  \bvol{39}~(18),  \pg{6273--6281}.

\bibitem[Baron {\em et~al.\/}(2008)Baron, B{\l}awzdziewicz \& Wajnryb]{baron08}
{\sc \au{Baron, M.}, \au{B{\l}awzdziewicz, J.} \& \au{Wajnryb, E.}} \yr{2008}
  \at{{Hydrodynamic crystals: Collective dynamics of regular arrays of
  spherical particles in a parallel-wall channel}}.  \jt{Phys. Rev. Lett.}
  \bvol{100}~(17),  \pg{174502}.

\bibitem[Bechinger {\em et~al.\/}(2016)Bechinger, Di~Leonardo, L{\"o}wen,
  Reichhardt, Volpe \& Volpe]{bechinger16}
{\sc \au{Bechinger, C.}, \au{Di~Leonardo, R.}, \au{L{\"o}wen, H.},
  \au{Reichhardt, C.}, \au{Volpe, G.} \& \au{Volpe, G.}} \yr{2016}  \at{Active
  particles in complex and crowded environments}.  \jt{Rev. Mod. Phys.}
  \bvol{88}~(4),  \pg{045006}.

\bibitem[Becker \& Braack(2001)]{BeckerBraack2001}
{\sc \au{Becker, R.} \& \au{Braack, M.}} \yr{2001}  \at{A finite element
  pressure gradient stabilization for the {S}tokes equations based on local
  projections}.  \jt{Calcolo}  \bvol{38}~(4),  \pg{173--199}.

\bibitem[Benesch {\em et~al.\/}(2003)Benesch, Yiacoumi \& Tsouris]{benesch03}
{\sc \au{Benesch, T.}, \au{Yiacoumi, S.} \& \au{Tsouris, C.}} \yr{2003}
  \at{Brownian motion in confinement}.  \jt{Phys. Rev. E}  \bvol{68}~(2),
  \pg{021401}.

\bibitem[Bhattacharya(2008)]{bhattacharya08}
{\sc \au{Bhattacharya, S.}} \yr{2008}  \at{Cooperative motion of spheres
  arranged in periodic grids between two parallel walls}.  \jt{J. Chem. Phys.}
  \bvol{128}~(7),  \pg{074709}.

\bibitem[Bhattacharya \& B{\l}awzdziewicz(2002)]{bhattacharya02}
{\sc \au{Bhattacharya, S.} \& \au{B{\l}awzdziewicz, J.}} \yr{2002}  \at{Image
  system for {S}tokes-flow singularity between two parallel planar walls}.
  \jt{J. Math. Phys.}  \bvol{43}~(11),  \pg{5720--5731}.

\bibitem[Bhattacharya {\em et~al.\/}(2005)Bhattacharya, B{\l}awzdziewicz \&
  Wajnryb]{bhattacharya05b}
{\sc \au{Bhattacharya, S.}, \au{B{\l}awzdziewicz, J.} \& \au{Wajnryb, E.}}
  \yr{2005}  \at{Hydrodynamic interactions of spherical particles in
  suspensions confined between two planar walls}.  \jt{J. Fluid Mech.}
  \bvol{541},  \pg{263--292}.

\bibitem[Bhattacharya {\em et~al.\/}(2006)Bhattacharya, B{\l}awzdziewicz \&
  Wajnryb]{bhattacharya06pof}
{\sc \au{Bhattacharya, S.}, \au{B{\l}awzdziewicz, J.} \& \au{Wajnryb, E.}}
  \yr{2006}  \at{{Hydrodynamic interactions of spherical particles in
  Poiseuille flow between two parallel walls}}.  \jt{Phys. Fluids}
  \bvol{18}~(5),  \pg{053301}.

\bibitem[Bilbao {\em et~al.\/}(2013)Bilbao, Wajnryb, Vanapalli \&
  B{\l}awzdziewicz]{bilbao13}
{\sc \au{Bilbao, A.}, \au{Wajnryb, E.}, \au{Vanapalli, S.~A.} \&
  \au{B{\l}awzdziewicz, J.}} \yr{2013}  \at{Nematode locomotion in unconfined
  and confined fluids}.  \jt{Phys. Fluids}  \bvol{25}~(8),  \pg{081902}.

\bibitem[Blake(1971)]{blake71}
{\sc \au{Blake, J.~R.}} \yr{1971}  \at{A note on the image system for a
  {S}tokeslet in a no-slip boundary}.  \jt{Math. Proc. Camb. Phil. Soc.}
  \bvol{70}~(02),  \pg{303--310}.

\bibitem[B{\l}awzdziewicz \& Wajnryb(2008)]{blaw08}
{\sc \au{B{\l}awzdziewicz, J.} \& \au{Wajnryb, E.}} \yr{2008}  \at{{An analysis
  of the far-field response to external forcing of a suspension in the Stokes
  flow in a parallel-wall channel}}.  \jt{Phys. Fluids}  \bvol{20}~(9),
  \pg{093303}.

\bibitem[Braack \& Richter(2006)]{BraackRichter2006d}
{\sc \au{Braack, M.} \& \au{Richter, T.}} \yr{2006}  \at{Solutions of {3D}
  {N}avier-{S}tokes benchmark problems with adaptive finite elements}.
  \jt{Comput. Fluids}  \bvol{35}~(4),  \pg{372--392}.

\bibitem[Bracewell(1999)]{bracewell99}
{\sc \au{Bracewell, R.}} \yr{1999} {\em {The Fourier Transform and Its
  Applications}\/}.  \publ{McGraw-Hill, Pennsylvania, U.S.A.}

\bibitem[Brenner(1999)]{brenner99}
{\sc \au{Brenner, M.~P.}} \yr{1999}  \at{Screening mechanisms in
  sedimentation}.  \jt{Phys. Fluids}  \bvol{11}~(4),  \pg{754--772}.

\bibitem[Brotto {\em et~al.\/}(2013)Brotto, Caussin, Lauga \&
  Bartolo]{brotto13}
{\sc \au{Brotto, T.}, \au{Caussin, J.-B.}, \au{Lauga, E.} \& \au{Bartolo, D.}}
  \yr{2013}  \at{Hydrodynamics of confined active fluids}.  \jt{Phys. Rev.
  Lett.}  \bvol{110}~(3),  \pg{038101}.

\bibitem[Campbell {\em et~al.\/}(2004)Campbell, Wilkinson, Manz, Camilleri \&
  Humphreys]{campbell04}
{\sc \au{Campbell, L.~C.}, \au{Wilkinson, M.~J.}, \au{Manz, A.}, \au{Camilleri,
  P.} \& \au{Humphreys, C.~J.}} \yr{2004}  \at{{Electrophoretic manipulation of
  single DNA molecules in nanofabricated capillaries}}.  \jt{Lab Chip}
  \bvol{4}~(3),  \pg{225--229}.

\bibitem[Copson(1961)]{copson61}
{\sc \au{Copson, E.~T.}} \yr{1961}  \at{On certain dual integral equations}.
  \jt{Glasgow Math. J.}  \bvol{5}~(1),  \pg{21--24}.

\bibitem[Cross {\em et~al.\/}(2007)Cross, Strychalski \& Craighead]{cross07}
{\sc \au{Cross, J.~D.}, \au{Strychalski, E.~A.} \& \au{Craighead, H.~G.}}
  \yr{2007}  \at{{Size-dependent DNA mobility in nanochannels}}.  \jt{J. Appl.
  Phys.}  \bvol{102}~(2),  \pg{024701}.

\bibitem[Daddi-Moussa-Ider \& Gekle(2016)]{daddi16c}
{\sc \au{Daddi-Moussa-Ider, A.} \& \au{Gekle, S.}} \yr{2016}  \at{Hydrodynamic
  interaction between particles near elastic interfaces}.  \jt{J. Chem. Phys.}
  \bvol{145}~(1),  \pg{014905}.

\bibitem[Daddi-Moussa-Ider \& Gekle(2017)]{daddi17b}
{\sc \au{Daddi-Moussa-Ider, A.} \& \au{Gekle, S.}} \yr{2017}  \at{Hydrodynamic
  mobility of a solid particle near a spherical elastic membrane: Axisymmetric
  motion}.  \jt{Phys. Rev. E}  \bvol{95},  \pg{013108}.

\bibitem[Daddi-Moussa-Ider \& Gekle(2018)]{daddi2018brownian}
{\sc \au{Daddi-Moussa-Ider, A.} \& \au{Gekle, S.}} \yr{2018}  \at{{Brownian
  motion near an elastic cell membrane: A theoretical study}}.  \jt{Eur. Phys.
  J. E}  \bvol{41}~(2),  \pg{19}.

\bibitem[Daddi-Moussa-Ider {\em et~al.\/}(2016)Daddi-Moussa-Ider, Guckenberger
  \& Gekle]{daddi16b}
{\sc \au{Daddi-Moussa-Ider, A.}, \au{Guckenberger, A.} \& \au{Gekle, S.}}
  \yr{2016}  \at{Particle mobility between two planar elastic membranes:
  {B}rownian motion and membrane deformation}.  \jt{Phys. Fluids}
  \bvol{28}~(7),  \pg{071903}.

\bibitem[Daddi-Moussa-Ider {\em et~al.\/}(2019)Daddi-Moussa-Ider, Kaoui \&
  L{\"o}wen]{daddi19jpsj}
{\sc \au{Daddi-Moussa-Ider, A.}, \au{Kaoui, B.} \& \au{L{\"o}wen, H.}}
  \yr{2019}  \at{{Axisymmetric flow due to a Stokeslet near a finite-sized
  elastic membrane}}.  \jt{J. Phys. Soc. Jpn.}  \bvol{88}~(5),  \pg{054401}.

\bibitem[Daddi-Moussa-Ider {\em et~al.\/}(2020)Daddi-Moussa-Ider, Lisicki,
  L{\"o}wen \& Menzel]{daddi-compo20}
{\sc \au{Daddi-Moussa-Ider, A.}, \au{Lisicki, M.}, \au{L{\"o}wen, H.} \&
  \au{Menzel, A.~M.}} \yr{2020}  \at{Dynamics of a microswimmer--microplatelet
  composite}.  \jt{Phys. Fluids}  \bvol{32}~(2),  \pg{021902}.

\bibitem[Daddi-Moussa-Ider {\em et~al.\/}(2018)Daddi-Moussa-Ider, Lisicki,
  Mathijssen, Hoell, Goh, B{\l}awzdziewicz, Menzel \&
  L{\"o}wen]{daddi2018state}
{\sc \au{Daddi-Moussa-Ider, A.}, \au{Lisicki, M.}, \au{Mathijssen, A. J.
  T.~M.}, \au{Hoell, C.}, \au{Goh, S.}, \au{B{\l}awzdziewicz, J.}, \au{Menzel,
  A.~M.} \& \au{L{\"o}wen, H.}} \yr{2018}  \at{State diagram of a three-sphere
  microswimmer in a channel}.  \jt{J. Phys.: Condens. Matter}  \bvol{30}~(25),
  \pg{254004}.

\bibitem[Dai {\em et~al.\/}(2013)Dai, Tree, van~der Maarel, Dorfman \&
  Doyle]{dai13}
{\sc \au{Dai, L.}, \au{Tree, D.~R.}, \au{van~der Maarel, J. R.~C.},
  \au{Dorfman, K.~D.} \& \au{Doyle, P.~S.}} \yr{2013}  \at{Revisiting blob
  theory for {DNA} diffusivity in slitlike confinement}.  \jt{Phys. Rev. Lett.}
   \bvol{110},  \pg{168105}.

\bibitem[Daiguji {\em et~al.\/}(2004)Daiguji, Yang, Szeri \&
  Majumdar]{daiguji04}
{\sc \au{Daiguji, H.}, \au{Yang, P.}, \au{Szeri, A.~J.} \& \au{Majumdar, A.}}
  \yr{2004}  \at{Electrochemomechanical energy conversion in nanofluidic
  channels}.  \jt{Nano Lett.}  \bvol{4}~(12),  \pg{2315--2321}.

\bibitem[Dalal {\em et~al.\/}(2020)Dalal, Farutin \& Misbah]{dalal20}
{\sc \au{Dalal, S.}, \au{Farutin, A.} \& \au{Misbah, C.}} \yr{2020}
  \at{Amoeboid swimming in compliant channel}.  \jt{Soft Matter}  \bvol{16},
  \pg{1599--1613}.

\bibitem[Darwiche {\em et~al.\/}(2013)Darwiche, Ingremeau, Amarouchene, Maali,
  Dufour \& Kellay]{darwiche13}
{\sc \au{Darwiche, A.}, \au{Ingremeau, F.}, \au{Amarouchene, Y.}, \au{Maali,
  A.}, \au{Dufour, I.} \& \au{Kellay, H.}} \yr{2013}  \at{Rheology of polymer
  solutions using colloidal-probe atomic force microscopy}.  \jt{Phys. Rev. E}
  \bvol{87}~(6),  \pg{062601}.

\bibitem[Davis \& Rabinowitz(2007)]{davis07}
{\sc \au{Davis, P.~J.} \& \au{Rabinowitz, P.}} \yr{2007} {\em {Methods of
  Numerical Integration}\/}.  \publ{Courier Corporation, North Chelmsford,
  Massachusetts, U.S.A.}

\bibitem[Doyle {\em et~al.\/}(2002)Doyle, Bibette, Bancaud \& Viovy]{doyle02}
{\sc \au{Doyle, P.~S.}, \au{Bibette, J.}, \au{Bancaud, A.} \& \au{Viovy,
  J.-L.}} \yr{2002}  \at{{Self-assembled magnetic matrices for DNA separation
  chips}}.  \jt{Science}  \bvol{295}~(5563),  \pg{2237--2237}.

\bibitem[Driscoll \& Delmotte(2019)]{driscoll19}
{\sc \au{Driscoll, M.} \& \au{Delmotte, B.}} \yr{2019}  \at{Leveraging
  collective effects in externally driven colloidal suspensions: Experiments
  and simulations}.  \jt{Curr. Opin. Colloid Interface Sci.}  \bvol{40},
  \pg{42--57}.

\bibitem[Dufour {\em et~al.\/}(2012)Dufour, Maali, Amarouchene, Ayela,
  Caillard, Darwiche, Guirardel, Kellay, Lemaire, Mathieu {\em
  et~al.\/}]{dufour12}
{\sc \au{Dufour, I.}, \au{Maali, A.}, \au{Amarouchene, Y.}, \au{Ayela, C.},
  \au{Caillard, B.}, \au{Darwiche, A.}, \au{Guirardel, M.}, \au{Kellay, H.},
  \au{Lemaire, E.}, \au{Mathieu, F.} \& \au{others}} \yr{2012}  \at{{The
  microcantilever: A versatile tool for measuring the rheological properties of
  complex fluids}}.  \jt{J. Sens.}  \bvol{2012}.

\bibitem[Dufresne {\em et~al.\/}(2001)Dufresne, Altman \& Grier]{dufresne01}
{\sc \au{Dufresne, E.~R.}, \au{Altman, D.} \& \au{Grier, D.~G.}} \yr{2001}
  \at{{B}rownian dynamics of a sphere between parallel walls}.  \jt{Europhys.
  Lett.}  \bvol{53}~(2),  \pg{264}.

\bibitem[Faucheux \& Libchaber(1994)]{faucheux94}
{\sc \au{Faucheux, L.~P.} \& \au{Libchaber, A.~J.}} \yr{1994}  \at{Confined
  {B}rownian motion}.  \jt{Phys. Rev. E}  \bvol{49},  \pg{5158--5163}.

\bibitem[Fax\'{e}n(1921)]{faxen21}
{\sc \au{Fax\'{e}n, H.}} \yr{1921}  \at{Einwirkung der {Gef\"{a}ssw\"{a}nde}
  auf den {Widerstand} gegen die {Bewegung} einer kleinen {Kugel} in einer
  z\"{a}hen {Fl\"{u}ssigkeit}}. PhD thesis, Uppsala University, Uppsala,
  Sweden.

\bibitem[Felderhof(2006)]{felderhof06twoWalls}
{\sc \au{Felderhof, B.~U.}} \yr{2006}  \at{Diffusion and velocity relaxation of
  a {B}rownian particle immersed in a viscous compressible fluid confined
  between two parallel plane walls}.  \jt{J. Chem. Phys.}  \bvol{124}~(5),
  \pg{054111}.

\bibitem[Felderhof(2010{\natexlab{{\em a\/}}})]{felderhof10echoing}
{\sc \au{Felderhof, B.~U.}} \yr{2010{\natexlab{{\em a\/}}}}  \at{Echoing in a
  viscous compressible fluid confined between two parallel plane walls}.
  \jt{J. Fluid Mech.}  \bvol{656},  \pg{223--230}.

\bibitem[Felderhof(2010{\natexlab{{\em b\/}}})]{felderhof10loss}
{\sc \au{Felderhof, B.~U.}} \yr{2010{\natexlab{{\em b\/}}}}  \at{Loss of
  momentum in a viscous compressible fluid due to no-slip boundary condition at
  one or two planar walls}.  \jt{J. Chem. Phys.}  \bvol{133}~(7),  \pg{074707}.

\bibitem[Fran{\c{c}}ois {\em et~al.\/}(2009)Fran{\c{c}}ois, Amarouchene, Lounis
  \& Kellay]{franccois09}
{\sc \au{Fran{\c{c}}ois, N.}, \au{Amarouchene, Y.}, \au{Lounis, B.} \&
  \au{Kellay, H.}} \yr{2009}  \at{Polymer conformations and hysteretic stresses
  in nonstationary flows of polymer solutions}.  \jt{Europhys. Lett.}
  \bvol{86}~(3),  \pg{34002}.

\bibitem[Fran{\c{c}}ois {\em et~al.\/}(2008)Fran{\c{c}}ois, Lasne, Amarouchene,
  Lounis \& Kellay]{franccois08}
{\sc \au{Fran{\c{c}}ois, N.}, \au{Lasne, D.}, \au{Amarouchene, Y.}, \au{Lounis,
  B.} \& \au{Kellay, H.}} \yr{2008}  \at{Drag enhancement with polymers}.
  \jt{Phys. Rev. Lett.}  \bvol{100}~(1),  \pg{018302}.

\bibitem[Ganatos {\em et~al.\/}(1980{\natexlab{{\em a\/}}})Ganatos, Pfeffer \&
  Weinbaum]{ganatos80b}
{\sc \au{Ganatos, P.}, \au{Pfeffer, R.} \& \au{Weinbaum, S.}}
  \yr{1980{\natexlab{{\em a\/}}}}  \at{A strong interaction theory for the
  creeping motion of a sphere between plane parallel boundaries. {P}art 2.
  {P}arallel motion}.  \jt{J. Fluid Mech.}  \bvol{99},  \pg{755--783}.

\bibitem[Ganatos {\em et~al.\/}(1980{\natexlab{{\em b\/}}})Ganatos, Weinbaum \&
  Pfeffer]{ganatos80a}
{\sc \au{Ganatos, P.}, \au{Weinbaum, S.} \& \au{Pfeffer, R.}}
  \yr{1980{\natexlab{{\em b\/}}}}  \at{A strong interaction theory for the
  creeping motion of a sphere between plane parallel boundaries. {P}art 1.
  {P}erpendicular motion}.  \jt{J. Fluid Mech.}  \bvol{99},  \pg{739--753}.

\bibitem[Gompper {\em et~al.\/}(2020)Gompper, Winkler, Speck, Solon, Nardini,
  Peruani, L{\"o}wen, Golestanian, Kaupp, Alvarez {\em et~al.\/}]{gompper20}
{\sc \au{Gompper, G.}, \au{Winkler, R.~G.}, \au{Speck, T.}, \au{Solon, A.},
  \au{Nardini, C.}, \au{Peruani, F.}, \au{L{\"o}wen, H.}, \au{Golestanian, R.},
  \au{Kaupp, U.~B.}, \au{Alvarez, L.} \& \au{others}} \yr{2020}  \at{The 2020
  motile active matter roadmap}.  \jt{J. Phys.: Condens. Matter}
  \bvol{32}~(19),  \pg{193001}.

\bibitem[Graham(2011)]{graham11}
{\sc \au{Graham, M.~D.}} \yr{2011}  \at{Fluid dynamics of dissolved polymer
  molecules in confined geometries}.  \jt{Ann. Rev. Fluid Mech.}  \bvol{43},
  \pg{273--298}.

\bibitem[Griggs {\em et~al.\/}(2007)Griggs, Zinchenko \& Davis]{griggs07}
{\sc \au{Griggs, A.~J.}, \au{Zinchenko, A.~Z.} \& \au{Davis, R.~H.}} \yr{2007}
  \at{{Low-Reynolds-number motion of a deformable drop between two parallel
  plane walls}}.  \jt{Int. J. Multiph. Flow}  \bvol{33}~(2),  \pg{182--206}.

\bibitem[Hackborn(1990)]{hackborn90}
{\sc \au{Hackborn, W.~W.}} \yr{1990}  \at{{Asymmetric Stokes flow between
  parallel planes due to a rotlet}}.  \jt{J. Fluid Mech.}  \bvol{218},
  \pg{531--546}.

\bibitem[Happel \& Brenner(1983)]{happel12}
{\sc \au{Happel, J.} \& \au{Brenner, H.}} \yr{1983} {\em {Low Reynolds Number
  Hydrodynamics: With Special Applications to Particulate Media}\/}.
  \publ{Springer Netherlands, Martinus Nijhoff Publishers, The Hague, The
  Netherlands}.

\bibitem[Imai(1973)]{imai73}
{\sc \au{Imai, I.}} \yr{1973} {\em {Fluid Dynamics (Ry\={u}tai Rikigaku)}\/}.
  \publ{Syokabo Publishing, Tokyo, Japan [in Japanese]}.

\bibitem[Janssen \& Anderson(2007)]{janssen07}
{\sc \au{Janssen, P. J.~A.} \& \au{Anderson, P.~D.}} \yr{2007}
  \at{Boundary-integral method for drop deformation between parallel plates}.
  \jt{Phys. Fluids}  \bvol{19}~(4),  \pg{043602}.

\bibitem[Janssen \& Anderson(2008)]{janssen08}
{\sc \au{Janssen, P. J.~A.} \& \au{Anderson, P.~D.}} \yr{2008}  \at{A
  boundary-integral model for drop deformation between two parallel plates with
  non-unit viscosity ratio drops}.  \jt{J. Comp. Phys.}  \bvol{227}~(20),
  \pg{8807--8819}.

\bibitem[Jones {\em et~al.\/}(2013)Jones, van~der Maarel \& Doyle]{jones13}
{\sc \au{Jones, J.~J.}, \au{van~der Maarel, J. R.~C.} \& \au{Doyle, P.~S.}}
  \yr{2013}  \at{{Intrachain dynamics of large ds{DNA} confined to slitlike
  channels}}.  \jt{Phys. Rev. Lett.}  \bvol{110}~(6),  \pg{068101}.

\bibitem[Jones(2004)]{jones04}
{\sc \au{Jones, R.~B.}} \yr{2004}  \at{{Spherical particle in Poiseuille flow
  between planar walls}}.  \jt{J. Chem. Phys.}  \bvol{121}~(1),  \pg{483--500}.

\bibitem[Kim(1983)]{kim83}
{\sc \au{Kim, M.~U.}} \yr{1983}  \at{{Axisymmetric Stokes flow due to a point
  force near a circular disk}}.  \jt{J. Phys. Soc. Jpn.}  \bvol{52}~(2),
  \pg{449--455}.

\bibitem[Kim \& Karrila(2013)]{kim13}
{\sc \au{Kim, S.} \& \au{Karrila, S.~J.}} \yr{2013} {\em {Microhydrodynamics:
  Principles and Selected Applications}\/}.  \publ{Courier Corporation, North
  Chelmsford, Massachusetts, U.S.A.}

\bibitem[Kushch(2013)]{kushch13}
{\sc \au{Kushch, V.~I.}} \yr{2013} {\em Micromechanics of Composites: Multipole
  Expansion Approach\/}.  \publ{Butterworth-Heinemann, Oxford, U.K.}

\bibitem[Kushch \& Sangani(2000)]{kushch00}
{\sc \au{Kushch, V.~I.} \& \au{Sangani, A.~S.}} \yr{2000}  \at{Conductivity of
  a composite containing uniformly oriented penny--shaped cracks or perfectly
  conducting discs}.  \jt{Proc. Roy. Soc. A.: Math. Phys.}  \bvol{456}~(1995),
  \pg{683--699}.

\bibitem[Lasne {\em et~al.\/}(2008)Lasne, Maali, Amarouchene, Cognet, Lounis \&
  Kellay]{lasne2008velocity}
{\sc \au{Lasne, D.}, \au{Maali, A.}, \au{Amarouchene, Y.}, \au{Cognet, L.},
  \au{Lounis, B.} \& \au{Kellay, H.}} \yr{2008}  \at{Velocity profiles of water
  flowing past solid glass surfaces using fluorescent nanoparticles and
  molecules as velocity probes}.  \jt{Phys. Rev. Lett.}  \bvol{100}~(21),
  \pg{214502}.

\bibitem[Lauga(2016)]{lauga2016ARFM}
{\sc \au{Lauga, E.}} \yr{2016}  \at{Bacterial hydrodynamics}.  \jt{Ann. Rev.
  Fluid Mech.}  \bvol{48},  \pg{105--130}.

\bibitem[Lauga {\em et~al.\/}(2007)Lauga, Brenner \& Stone]{lauga07noslip}
{\sc \au{Lauga, E.}, \au{Brenner, M.} \& \au{Stone, H.}} \yr{2007}
  \at{Microfluidics: {T}he no-slip boundary condition}.  \bt{In {\em {Springer
  Handbook of Experimental Fluid Mechanics}\/}},  \pg{pp. 1219--1240}.
  \publ{Springer}.

\bibitem[Lauga \& Powers(2009)]{lauga09}
{\sc \au{Lauga, E.} \& \au{Powers, T.~R.}} \yr{2009}  \at{The hydrodynamics of
  swimming microorganisms}.  \jt{Rep. Prog. Phys.}  \bvol{72}~(9),
  \pg{096601}.

\bibitem[Lauga \& Squires(2005)]{lauga05}
{\sc \au{Lauga, E.} \& \au{Squires, T.~M.}} \yr{2005}  \at{Brownian motion near
  a partial-slip boundary: A local probe of the no-slip condition}.  \jt{Phys.
  Fluids}  \bvol{17}~(10),  \pg{103102}.

\bibitem[Le~Goff {\em et~al.\/}(2017)Le~Goff, Kaoui, Kurzawa, Haszon \&
  Salsac]{legoff17}
{\sc \au{Le~Goff, A.}, \au{Kaoui, B.}, \au{Kurzawa, G.}, \au{Haszon, B.} \&
  \au{Salsac, A.-V.}} \yr{2017}  \at{Squeezing bio-capsules into a
  constriction: deformation till break-up}.  \jt{Soft Matter}  \bvol{13}~(41),
  \pg{7644--7648}.

\bibitem[Leal(1980)]{leal80}
{\sc \au{Leal, L.~G.}} \yr{1980}  \at{Particle motions in a viscous fluid}.
  \jt{Ann. Rev. Fluid Mech.}  \bvol{12}~(1),  \pg{435--476}.

\bibitem[Lee {\em et~al.\/}(1979)Lee, Chadwick \& Leal]{lee79}
{\sc \au{Lee, S.~H.}, \au{Chadwick, R.~S.} \& \au{Leal, L.~G.}} \yr{1979}
  \at{{Motion of a sphere in the presence of a plane interface. Part 1. An
  approximate solution by generalization of the method of Lorentz}}.  \jt{J.
  Fluid Mech.}  \bvol{93},  \pg{705--726}.

\bibitem[Lee \& Leal(1980)]{lee80}
{\sc \au{Lee, S.~H.} \& \au{Leal, L.~G.}} \yr{1980}  \at{{Motion of a sphere in
  the presence of a plane interface. Part 2. An exact solution in bipolar
  co-ordinates}}.  \jt{J. Fluid Mech.}  \bvol{98},  \pg{193--224}.

\bibitem[Lin {\em et~al.\/}(2000)Lin, Yu \& Rice]{lin00}
{\sc \au{Lin, B.}, \au{Yu, J.} \& \au{Rice, S.~A.}} \yr{2000}  \at{Direct
  measurements of constrained {B}rownian motion of an isolated sphere between
  two walls}.  \jt{Phys. Rev. E}  \bvol{62},  \pg{3909--3919}.

\bibitem[Liron(1978)]{liron78cilia}
{\sc \au{Liron, N.}} \yr{1978}  \at{Fluid transport by cilia between parallel
  plates}.  \jt{J. Fluid Mech.}  \bvol{86}~(4),  \pg{705--726}.

\bibitem[Liron \& Blake(1981)]{liron81}
{\sc \au{Liron, N.} \& \au{Blake, J.~R.}} \yr{1981}  \at{Existence of viscous
  eddies near boundaries}.  \jt{J. Fluid Mech.}  \bvol{107},  \pg{109--129}.

\bibitem[Liron \& Mochon(1976)]{liron76}
{\sc \au{Liron, N.} \& \au{Mochon, S.}} \yr{1976}  \at{Stokes flow for a
  {S}tokeslet between two parallel flat plates}.  \jt{J. Eng. Math.}
  \bvol{10}~(4),  \pg{287--303}.

\bibitem[Lobry \& Ostrowsky(1996)]{lobry96}
{\sc \au{Lobry, L.} \& \au{Ostrowsky, N.}} \yr{1996}  \at{Diffusion of
  {B}rownian particles trapped between two walls: Theory and
  dynamic-light-scattering measurements}.  \jt{Phys. Rev. B}  \bvol{53},
  \pg{12050--12056}.

\bibitem[Lorentz(1907)]{lorentz07}
{\sc \au{Lorentz, H.~A.}} \yr{1907}  \at{{Ein allgemeiner Satz, die Bewegung
  einer reibenden Fl\"{u}ssigkeit betreffend, nebst einigen Anwendungen
  desselben}}.  \jt{Abh. Theor. Phys.}  \bvol{1},  \pg{23}.

\bibitem[Marini Bettolo~Marconi \& Melchionna(2012)]{marini12}
{\sc \au{Marini Bettolo~Marconi, U.} \& \au{Melchionna, S.}} \yr{2012}
  \at{{Charge transport in nanochannels: A molecular theory}}.  \jt{Langmuir}
  \bvol{28}~(38),  \pg{13727--13740}.

\bibitem[Mathijssen {\em et~al.\/}(2016)Mathijssen, Doostmohammadi, Yeomans \&
  Shendruk]{mathijssen2015hydrodynamics}
{\sc \au{Mathijssen, A. J. T.~M.}, \au{Doostmohammadi, A.}, \au{Yeomans, J.~M.}
  \& \au{Shendruk, T.~N.}} \yr{2016}  \at{Hydrodynamics of microswimmers in
  films}.  \jt{J. Fluid Mech.}  \bvol{806},  \pg{35--70}.

\bibitem[Menzel(2013)]{menzel13}
{\sc \au{Menzel, A.~M.}} \yr{2013}  \at{Unidirectional laning and migrating
  cluster crystals in confined self-propelled particle systems}.  \jt{J. Phys.:
  Condens. Matter}  \bvol{25}~(50),  \pg{505103}.

\bibitem[Menzel(2015)]{menzel15}
{\sc \au{Menzel, A.~M.}} \yr{2015}  \at{Tuned, driven, and active soft matter}.
   \jt{Phys. Rep.}  \bvol{554},  \pg{1--45}.

\bibitem[Mewis \& Wagner(2012)]{mewis12}
{\sc \au{Mewis, J.} \& \au{Wagner, N.~J.}} \yr{2012} {\em {Colloidal Suspension
  Rheology}\/}.  \publ{Cambridge University Press, Cambridge, U.K.}

\bibitem[Miyazaki(1984)]{miyazaki84}
{\sc \au{Miyazaki, T.}} \yr{1984}  \at{The effect of a circular disk on the
  motion of a small particle in a viscous fluid}.  \jt{J. Phys. Soc. Jpn.}
  \bvol{53}~(3),  \pg{1017--1025}.

\bibitem[Moffatt(1964)]{moffatt64}
{\sc \au{Moffatt, H~Keith}} \yr{1964}  \at{Viscous and resistive eddies near a
  sharp corner}.  \jt{J. Fluid Mech.}  \bvol{18}~(1),  \pg{1--18}.

\bibitem[Oseen(1928)]{oseen28}
{\sc \au{Oseen, C.~W.}} \yr{1928} {\em Neuere {Methoden} und {Ergebnisse} in
  der {Hydrodynamik}\/}.  \publ{Leipzig, Akademische Verlagsgesellschaft, M. B.
  H., Germany}.

\bibitem[Ostapenko {\em et~al.\/}(2018)Ostapenko, Schwarzendahl, B\"oddeker,
  Kreis, Cammann, Mazza \& B\"aumchen]{ostapenko2018}
{\sc \au{Ostapenko, T.}, \au{Schwarzendahl, F.~J.}, \au{B\"oddeker, T.~J.},
  \au{Kreis, C.~T.}, \au{Cammann, J.}, \au{Mazza, M.~G.} \& \au{B\"aumchen,
  O.}} \yr{2018}  \at{Curvature-guided motility of microalgae in geometric
  confinement}.  \jt{Phys. Rev. Lett.}  \bvol{120},  \pg{068002}.

\bibitem[Ozarkar \& Sangani(2008)]{ozarkar08}
{\sc \au{Ozarkar, S.~S.} \& \au{Sangani, A.~S.}} \yr{2008}  \at{{A method for
  determining Stokes flow around particles near a wall or in a thin film
  bounded by a wall and a gas-liquid interface}}.  \jt{Phys. Fluids}
  \bvol{20}~(6),  \pg{063301}.

\bibitem[Park \& Dimitrakopoulos(2013)]{park13}
{\sc \au{Park, S.-Y.} \& \au{Dimitrakopoulos, P.}} \yr{2013}  \at{Transient
  dynamics of an elastic capsule in a microfluidic constriction}.  \jt{Soft
  Matter}  \bvol{9}~(37),  \pg{8844--8855}.

\bibitem[Persson \& Tegenfeldt(2010)]{persson10}
{\sc \au{Persson, F.} \& \au{Tegenfeldt, J.~O.}} \yr{2010}  \at{{DNA in
  nanochannels—directly visualizing genomic information}}.  \jt{Chem. Soc.
  Rev.}  \bvol{39}~(3),  \pg{985--999}.

\bibitem[Polyanin \& Manzhirov(1998)]{polyanin98}
{\sc \au{Polyanin, A.~D.} \& \au{Manzhirov, A.~V.}} \yr{1998} {\em {Handbook of
  Integral Equations}\/}.  \publ{CRC Press, Boca Raton, Florida, U.S.A.}

\bibitem[Probstein(2005)]{probstein05}
{\sc \au{Probstein, R.~F.}} \yr{2005} {\em {Physicochemical Hydrodynamics: An
  Introduction}\/}.  \publ{John Wiley \& Sons, Hoboken, New Jersey, U.S.A.}

\bibitem[Reisner {\em et~al.\/}(2005)Reisner, Morton, Riehn, Wang, Yu, Rosen,
  Sturm, Chou, Frey \& Austin]{reisner05}
{\sc \au{Reisner, W.}, \au{Morton, K.~J.}, \au{Riehn, R.}, \au{Wang, Y.~M.},
  \au{Yu, Z.}, \au{Rosen, M.}, \au{Sturm, J.~C.}, \au{Chou, S.~Y.}, \au{Frey,
  E.} \& \au{Austin, R.~H.}} \yr{2005}  \at{{Statics and dynamics of single DNA
  molecules confined in nanochannels}}.  \jt{Phys. Rev. Lett.}  \bvol{94}~(19),
   \pg{196101}.

\bibitem[Richter(2017)]{Richter2017}
{\sc \au{Richter, T.}} \yr{2017} {\em {Fluid-structure Interactions. Models,
  Analysis and Finite Elements}\/},  \st{Lecture notes in computational science
  and engineering},  \vol{vol. 118}.  \publ{Springer}.

\bibitem[Riehn {\em et~al.\/}(2005)Riehn, Lu, Wang, Lim, Cox \&
  Austin]{riehn05}
{\sc \au{Riehn, R.}, \au{Lu, M.}, \au{Wang, Y.-M.}, \au{Lim, S.~F.}, \au{Cox,
  E.~C.} \& \au{Austin, R.~H.}} \yr{2005}  \at{Restriction mapping in
  nanofluidic devices}.  \jt{Proc. Nat. Acad. Sci. U.S.A.}  \bvol{102}~(29),
  \pg{10012--10016}.

\bibitem[Roy(2010)]{roy10}
{\sc \au{Roy, C.~J.}} \yr{2010} Review of discretization error estimators in
  scientific computing.  \bt{In {\em 48th AIAA Aerospace Sciences Meeting
  Including the New Horizons Forum and Aerospace Exposition\/}},  \pg{p. 126}.

\bibitem[Saintillan {\em et~al.\/}(2006)Saintillan, Shaqfeh \&
  Darve]{saintillan06}
{\sc \au{Saintillan, D.}, \au{Shaqfeh, E. S.~G.} \& \au{Darve, E.}} \yr{2006}
  \at{Effect of flexibility on the shear-induced migration of short-chain
  polymers in parabolic channel flow}.  \jt{J. Fluid Mech.}  \bvol{557},
  \pg{297--306}.

\bibitem[Shaebani {\em et~al.\/}(2020)Shaebani, Wysocki, Winkler, Gompper \&
  Rieger]{shaebani20}
{\sc \au{Shaebani, M.~R.}, \au{Wysocki, A.}, \au{Winkler, R.~G.}, \au{Gompper,
  G.} \& \au{Rieger, H.}} \yr{2020}  \at{Computational models for active
  matter}.  \jt{Nat. Rev. Phys.}  \pg{pp. 1--19}.

\bibitem[Smithies(1958)]{smithies58}
{\sc \au{Smithies, F.}} \yr{1958} {\em {Integral Equations}\/}.
  \publ{Cambridge University Press, Cambridge, U.K.}

\bibitem[Sneddon(1960)]{sneddon60}
{\sc \au{Sneddon, I.~N.}} \yr{1960}  \at{The elementary solution of dual
  integral equations}.  \jt{Glasgow Math. J.}  \bvol{4}~(3),  \pg{108--110}.

\bibitem[Sneddon(1966)]{sneddon66}
{\sc \au{Sneddon, I.~N.}} \yr{1966} {\em {Mixed Boundary Value Problems in
  Potential Theory}\/}.  \publ{North-Holland, Amsterdam, The Netherlands}.

\bibitem[Spiegel(1965)]{spiegel65}
{\sc \au{Spiegel, M.~R.}} \yr{1965} {\em {Laplace Transforms}\/}.
  \publ{McGraw-Hill, New York, U.S.A.}

\bibitem[Staben {\em et~al.\/}(2003)Staben, Zinchenko \& Davis]{staben03}
{\sc \au{Staben, M.~E.}, \au{Zinchenko, A.~Z.} \& \au{Davis, R.~H.}} \yr{2003}
  \at{Motion of a particle between two parallel plane walls in
  low-{R}eynolds-number {P}oiseuille flow}.  \jt{Phys. Fluids}  \bvol{15}~(6),
  \pg{1711--1733}.

\bibitem[Stein {\em et~al.\/}(2006)Stein, van~der Heyden, Koopmans \&
  Dekker]{stein06}
{\sc \au{Stein, D.}, \au{van~der Heyden, F. H.~J.}, \au{Koopmans, W. J.~A.} \&
  \au{Dekker, C.}} \yr{2006}  \at{{Pressure-driven transport of confined DNA
  polymers in fluidic channels}}.  \jt{Proc. Nat. Acad. Sci. U.S.A.}
  \bvol{103}~(43),  \pg{15853--15858}.

\bibitem[Stokes(1851)]{stokes51}
{\sc \au{Stokes, G.~G.}} \yr{1851}  \at{{On the effect of the internal friction
  of fluids on the motion of pendulums}}.  \jt{Trans. Cambridge Philos. Soc.}
  \bvol{9},  \pg{8}.

\bibitem[Strychalski {\em et~al.\/}(2008)Strychalski, Levy \&
  Craighead]{strychalski08}
{\sc \au{Strychalski, E.~A.}, \au{Levy, S.~L.} \& \au{Craighead, H.~G.}}
  \yr{2008}  \at{{Diffusion of DNA in nanoslits}}.  \jt{Macromolecules}
  \bvol{41}~(20),  \pg{7716--7721}.

\bibitem[Swan \& Brady(2007)]{swan07}
{\sc \au{Swan, J.~W.} \& \au{Brady, J.~F.}} \yr{2007}  \at{Simulation of
  hydrodynamically interacting particles near a no-slip boundary}.  \jt{Phys.
  Fluids}  \bvol{19}~(11),  \pg{113306}.

\bibitem[Swan \& Brady(2010)]{swan10}
{\sc \au{Swan, J.~W.} \& \au{Brady, J.~F.}} \yr{2010}  \at{Particle motion
  between parallel walls: Hydrodynamics and simulation}.  \jt{Phys. Fluids}
  \bvol{22}~(10),  \pg{103301}.

\bibitem[Swan \& Brady(2011)]{swan11}
{\sc \au{Swan, J.~W.} \& \au{Brady, J.~F.}} \yr{2011}  \at{The hydrodynamics of
  confined dispersions}.  \jt{J. Fluid Mech.}  \bvol{687},  \pg{254--299}.

\bibitem[Tang {\em et~al.\/}(2010)Tang, Levy, Trahan, Jones, Craighead \&
  Doyle]{tang10}
{\sc \au{Tang, J.}, \au{Levy, S.~L.}, \au{Trahan, D.~W.}, \au{Jones, J.~J.},
  \au{Craighead, H.~G.} \& \au{Doyle, P.~S.}} \yr{2010}  \at{Revisiting the
  conformation and dynamics of {DNA} in slitlike confinement}.
  \jt{Macromolecules}  \bvol{43}~(17),  \pg{7368--7377}.

\bibitem[Thakore \& Hickman(2015)]{thakore15}
{\sc \au{Thakore, V.} \& \au{Hickman, J.~J.}} \yr{2015}  \at{Charge relaxation
  dynamics of an electrolytic nanocapacitor}.  \jt{J. Phys. Chem. C}
  \bvol{119}~(4),  \pg{2121--2132}.

\bibitem[Tr{\"a}nkle {\em et~al.\/}(2016)Tr{\"a}nkle, Ruh \&
  Rohrbach]{traenkle16}
{\sc \au{Tr{\"a}nkle, B.}, \au{Ruh, D.} \& \au{Rohrbach, A.}} \yr{2016}
  \at{Interaction dynamics of two diffusing particles: contact times and
  influence of nearby surfaces}.  \jt{Soft Matter}  \bvol{12}~(10),
  \pg{2729--2736}.

\bibitem[Tr{\'e}gou{\"e}t {\em et~al.\/}(2018)Tr{\'e}gou{\"e}t, Salez, Monteux
  \& Reyssat]{tregouet18}
{\sc \au{Tr{\'e}gou{\"e}t, C.}, \au{Salez, T.}, \au{Monteux, C.} \&
  \au{Reyssat, M.}} \yr{2018}  \at{Transient deformation of a droplet near a
  microfluidic constriction: A quantitative analysis}.  \jt{Phys. Rev. Fluids}
  \bvol{3}~(5),  \pg{053603}.

\bibitem[Tr{\'e}gou{\"e}t {\em et~al.\/}(2019)Tr{\'e}gou{\"e}t, Salez, Monteux
  \& Reyssat]{tregouet19}
{\sc \au{Tr{\'e}gou{\"e}t, C.}, \au{Salez, T.}, \au{Monteux, C.} \&
  \au{Reyssat, M.}} \yr{2019}  \at{Microfluidic probing of the complex
  interfacial rheology of multilayer capsules}.  \jt{Soft Matter}
  \bvol{15}~(13),  \pg{2782--2790}.

\bibitem[Tricomi(1985)]{tricomi85}
{\sc \au{Tricomi, F.~G.}} \yr{1985} {\em {Integral Equations}\/}.
  \publ{Courier Corporation, Mineola, New York, U.S.A.}

\bibitem[Turner {\em et~al.\/}(1998)Turner, Perez, Lopez \&
  Craighead]{turner98}
{\sc \au{Turner, S.~W.}, \au{Perez, A.~M.}, \au{Lopez, A.} \& \au{Craighead,
  H.~G.}} \yr{1998}  \at{{Monolithic nanofluid sieving structures for DNA
  manipulation}}.  \jt{J. Vac. Sci. Technol. B}  \bvol{16}~(6),
  \pg{3835--3840}.

\bibitem[Uspal {\em et~al.\/}(2013)Uspal, Eral \& Doyle]{uspal13}
{\sc \au{Uspal, W.~E.}, \au{Eral, H.~B.} \& \au{Doyle, P.~S.}} \yr{2013}
  \at{Engineering particle trajectories in microfluidic flows using particle
  shape}.  \jt{Nat. Commun.}  \bvol{4},  \pg{2666}.

\bibitem[Widder(2015)]{widder15}
{\sc \au{Widder, D.~V.}} \yr{2015} {\em {Laplace Transform (PMS-6)}\/}.
  \publ{Princeton University Press, New Jersey, U.S.A}.

\bibitem[Wolfram(1999)]{wolfram99}
{\sc \au{Wolfram, S.}} \yr{1999} {\em The
  MATHEMATICA\textsuperscript{\tiny{\textregistered}} Book, Version 4\/}.
  \publ{Cambridge University Press, Cambridge, U.K.}

\bibitem[Wu {\em et~al.\/}(2016)Wu, Farutin, Hu, Thi{\'e}baud, Rafa{\"\i},
  Peyla, Lai \& Misbah]{wu16}
{\sc \au{Wu, H.}, \au{Farutin, A.}, \au{Hu, W.-F.}, \au{Thi{\'e}baud, M.},
  \au{Rafa{\"\i}, S.}, \au{Peyla, P.}, \au{Lai, M.-C.} \& \au{Misbah, C.}}
  \yr{2016}  \at{Amoeboid swimming in a channel}.  \jt{Soft Matter}
  \bvol{12}~(36),  \pg{7470--7484}.

\bibitem[Wu {\em et~al.\/}(2015)Wu, Thi{\'e}baud, Hu, Farutin, Rafai, Lai,
  Peyla \& Misbah]{wu15}
{\sc \au{Wu, H.}, \au{Thi{\'e}baud, M.}, \au{Hu, W.-F.}, \au{Farutin, A.},
  \au{Rafai, S.}, \au{Lai, M.-C.}, \au{Peyla, P.} \& \au{Misbah, C.}} \yr{2015}
   \at{Amoeboid motion in confined geometry}.  \jt{Phys. Rev. E}
  \bvol{92}~(5),  \pg{050701}.

\bibitem[Xia {\em et~al.\/}(2012)Xia, Yan \& Hou]{xia12}
{\sc \au{Xia, D.}, \au{Yan, J.} \& \au{Hou, S.}} \yr{2012}  \at{{Fabrication of
  nanofluidic biochips with nanochannels for applications in DNA analysis}}.
  \jt{Small}  \bvol{8}~(18),  \pg{2787--2801}.

\bibitem[Z{\"o}ttl \& Stark(2016)]{zottl16}
{\sc \au{Z{\"o}ttl, A.} \& \au{Stark, H.}} \yr{2016}  \at{Emergent behavior in
  active colloids}.  \jt{J. Phys.: Condens. Matter}  \bvol{28}~(25),
  \pg{253001}.

\end{thebibliography}


\end{document}